\documentclass[acmlarge, nonacm]{acmart}

\usepackage{xcolor} %
\usepackage{float}

\AtBeginDocument{%
  \providecommand\BibTeX{{%
    \normalfont B\kern-0.5em{\scshape i\kern-0.25em b}\kern-0.8em\TeX}}}

\setcopyright{acmcopyright}
\copyrightyear{2026}
\acmYear{2026}
\acmDOI{XXXXXXX.XXXXXXX}

\acmJournal{IMWUT}
\acmISBN{978-1-4503-XXXX-X/18/06}

\usepackage{xac}

\usepackage{textgreek}
\usepackage{longtable}
\usepackage{booktabs}
\usepackage{placeins} %

\begin{document}

\title{Voice-Based Conversational Agents as AI Care Assistants: Evaluating Personalization, Intervention Delivery Fidelity, and Privacy Expectations in Older Adults’ Care}

\author{Thalia Viranda}
\orcid{0009-0008-6933-3926}
\email{tv74@cornell.edu}
\affiliation{%
  \department{Information Science}
  \institution{Cornell Tech}
  \city{New York City}
  \state{NY}
  \country{USA}}

\author{Shupeng Wang}
\orcid{0009-0003-4106-5361}
\email{sw2456@cornell.edu}
\affiliation{%
  \institution{Cornell Tech}
  \city{New York City}
  \state{NY}
  \country{USA}}

\author{Jia Bo}
\orcid{0009-0004-6522-8345}
\email{jb2848@cornell.edu}
\affiliation{%
  \institution{Cornell Tech}
  \city{New York City}
  \state{NY}
  \country{USA}}

\author{Yuchen Yang}
\orcid{0009-0002-4148-7823}
\email{yy2458@cornell.edu}
\affiliation{%
  \institution{Cornell Tech}
  \city{New York City}
  \state{NY}
  \country{USA}}

\author{Hongjiao Zhang}
\orcid{0009-0003-4925-0879}
\affiliation{%
  \institution{Cornell Tech}
  \city{New York City}
  \state{NY}
  \country{USA}}

\author{Andrea Cuadra}
\orcid{0000-0002-1845-0240}
\email{acuadra@olin.edu}
\affiliation{%
  \department{Computer Science}
  \institution{Olin College}
  \city{Needham}
  \state{MA}
  \country{USA}}

\author{Daniel A. Adler}
\orcid{0000-0003-3328-0312}
\email{adlerdan@umich.edu}
\affiliation{%
  \institution{University of Michigan, Cornell University}
  \city{Ann Arbor}
  \state{MI}
  \country{USA}}

\author{Walter Boot}
\orcid{0000-0003-1047-5467}
\email{wab4001@med.cornell.edu}
\affiliation{%
  \department{Division of Geriatrics and Palliative Care, Center on Aging and Behavioral Research}
  \institution{Weill Cornell Medicine}
  \city{New York City}
  \state{NY}
  \country{USA}}

\author{Sara Czaja}
\orcid{0000-0002-8096-5413}
\email{sjc7004@med.cornell.edu}
\affiliation{%
  \department{Division of Geriatrics and Palliative Care, Center on Aging and Behavioral Research}
  \institution{Weill Cornell Medicine}
  \city{New York City}
  \state{NY}
  \country{USA}}

\author{Deborah Estrin}
\orcid{0000-0001-6477-0096}
\email{destrin@cornell.edu}
\affiliation{%
  \department{Computer Science}
  \institution{Cornell Tech}
  \city{New York City}
  \state{NY}
  \country{USA}}

\author{Tanzeem Choudhury}
\orcid{0000-0002-5952-4955}
\email{tanzeem.choudhury@cornell.edu}
\affiliation{%
  \department{Information Science}
  \institution{Cornell Tech}
  \city{New York City}
  \state{NY}
  \country{USA}}

\begin{abstract}
  Generative AI enables voice-based conversational agents to play more active roles in older adults’ care, including personalizing support and operating alongside clinicians and care partners. This shift creates two challenges: how to personalize support while faithfully delivering evidence-based interventions, and how to manage personal-data use and sharing across AI and care stakeholders. We designed and evaluated ARC, a voice-based conversational agent that operationalizes multidomain lifestyle intervention through sessions with predefined goals, participant context, and evidence-based components. In a 3-day remote study, 80 older adults completed 473 sessions across six session types (9,183 turns; averaging 19.4 turns and 6 minutes per session). Participants reported favorable usability, acceptability, and use intentions. System-level analyses showed strong domain-level but modest participant-specific contextual reuse, no significant personalization–fidelity association, and fidelity limitations centered on incomplete delivery of intervention components rather than off-protocol drift. Within care relationships, expectations for personal-data use and sharing varied by data type and recipient. Our findings support feasibility of conversational delivery of multidomain lifestyle intervention, offer methods and system-design insights for longitudinal evaluation, and motivate an evaluation agenda for generative AI health interventions: user-experience and outcome measures should be complemented by evidence of whether intended intervention processes were delivered as designed.

\end{abstract}



\begin{teaserfigure}
  \centering
    \includegraphics[
    width=0.83\textwidth,
    keepaspectratio
  ]{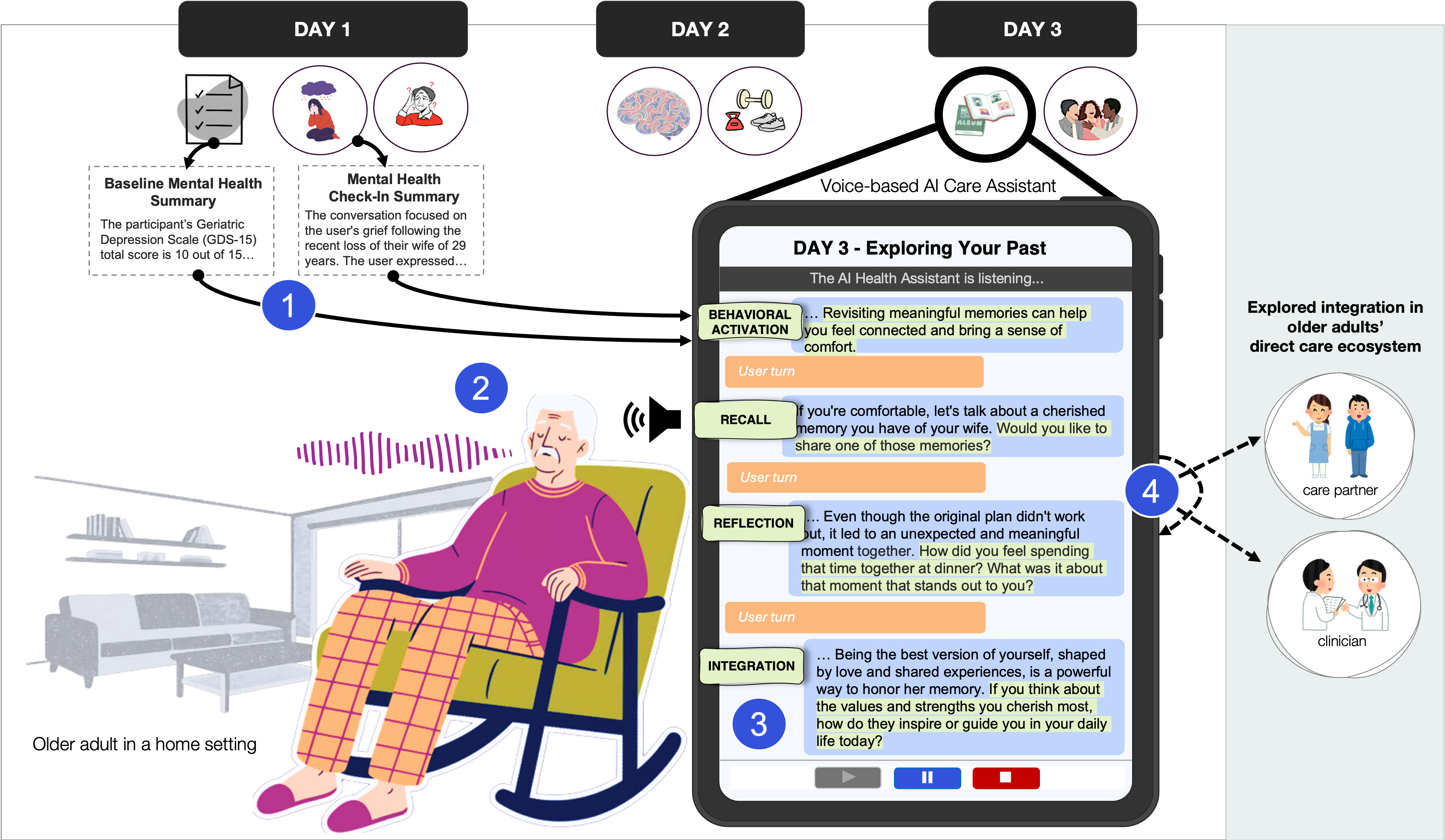}
  \caption{\textbf{ARC: Voice-Based AI Care Assistant in Use.} An older adult talks with ARC by voice at home during a Day~3 Exploring Your Past session. (1)~Baseline and prior-session summaries provide context; (2)~interaction is speech-to-speech; (3)~session protocols specify evidence-based processes, such as behavioral activation, recall, reflection, and integration; (4)~we explored older adults' expectations for personal-data use by ARC and sharing through ARC with clinicians and care partners.}
  \Description{An older adult in a rocking chair talks to a tablet showing an ARC session; labeled steps show personalization from prior summaries, voice interaction, evidence-based conversation steps, and possible sharing with a care partner and clinician.}
  \label{fig:teaser}
\end{teaserfigure}

\maketitle

\externaldocument{08_appendix}

\section{Introduction}
Despite rapid advances in voice-based conversational agents (VCAs) for health, little work has examined their use for delivering structured, evidence-based interventions for older adults with cognitive and mental health needs. Prior work has explored clinically informed voice-based interventions\cite{striegl_investigating_2022, wong_voice_2024, mathioudakis_voice_2026}, but has largely focused on conceptual design, usability, acceptability, or isolated behavioral support tasks. Existing systems primarily support information retrieval\cite{hoy_alexa_2018, google_google_2026}, open-ended interaction\cite{openai_introducing_2026, slingshot_ai_ash_2026, microsoft_copilot_2025}, health and well-being assessment\cite{ding_talktive_2022, jo_understanding_2023, kim_when_2026, myndyou_case_2021, mukherjee_polaris_2024}, or care communication and coordination\cite{yang_talk2care_2024, karlin_digital_2025, ellipsis_health_inc_meet_2025} (Figure \ref{fig:lit}), rather than delivering multi-step health and well-being interventions over time. What is missing is not another voice-based AI assistant or companion, but systems that translate  evidence-based interventions into generative conversations that adapt to individual context while preserving their intended intervention processes. We refer to systems that combine participant context with predefined intervention goals, required processes, and endpoints as \textit{\textbf{user-adaptive, therapeutically guided, closed-form}} systems. Such systems are particularly timely as population aging places growing demands on already constrained care systems.

Population aging is increasing demand for cognitive and behavioral health support, including for older adults at risk for or living with mild cognitive impairment (MCI)\cite{alzheimers_association_2026_2026, world_health_organization_decade_2020}. Approximately 10-15\% of older adults over 65 live with MCI, of whom roughly 10-40\% progress to dementia annually\cite{schmidtke_high_2008, petersen_mild_2009, irwin_predicting_2024}. At the same time, specialist care remains limited and family caregivers carry substantial long-term care responsibilities\cite{schulz_families_2016, institute_of_medicine_retooling_2008}. A growing body of evidence shows that \textit{multidomain lifestyle intervention}---a non-pharmacological approach targeting cognitive activity, physical activity, nutrition, and social engagement---can improve cognitive outcomes in older adults at risk for or living with MCI\cite{kivipelto_finnish_2013, kivipelto_world-wide_2020, baker_study_2024, salzman_associations_2022}. However, translating these structured, specialist-supported interventions into accessible digital delivery remains challenging, particularly for sustaining engagement\cite{castro_multi-domain_2023, kebede_digital_2022}, supporting multi-step protocols into usable interactions\cite{ de_souto_barreto_web-based_2021, mattli_digital_2025}, and maintaining fidelity while adapting to individual needs\cite{bott_face--face_2019, rao_delivering_2025}. Voice-based conversational agents (VCAs) are particularly well-suited to this delivery context. Spoken interaction can reduce the attentional, perceptual, and motor demands of screen-based tools\cite{pradhan_use_2020, brewer_empirical_2022}, while supporting brief interaction in naturalistic settings over weeks to months\cite{naylor_therapeutic_2008, berube_voice-based_2021, lima_discovering_2023, jansons_delivery_2022}. These affordances make voice a promising modality for longitudinal behavioral and cognitive support.

Realizing this potential requires moving beyond reactive or open-ended interaction toward personalized, therapeutically guided, closed-form VCAs. \textit{Personalization} in conversational agents refers to retaining, interpreting, and applying a user's history, preferences, and evolving needs to shape support across interactions\cite{hendrickx_take_2021}. At the same time, these systems must also maintain \textit{fidelity}, or adherence to the intervention processes intended to be delivered. Clinical intervention research describes this principle as \textit{flexibility within fidelity}\cite{fonagy_fidelity_2019, zilcha-mano_evidence-based_2022}. Personalization and fidelity are therefore complementary rather than opposing goals, as effective intervention delivery may depend on achieving both. For generative conversational agents, this raises a process-level evaluation question: \textit{how and to what extent can an agent's conversational behavior reflect participant context while maintaining fidelity to evidence-based intervention delivery?} 

Beyond intervention delivery within the human-AI interaction, using intelligent voice conversational agents to facilitate personalized health intervention also raises questions about how personal information should be used and shared across broader care relationships, where AI systems, clinicians, and care partners may have different roles and information needs\cite{frik_privacy_2019, hendrickx_take_2021, mynatt_technology_2025, kim_clinicians_2021}. Understanding these context-dependent expectations is important for designing appropriate information flows as such conversational interventions become embedded in care ecosystems.

To examine these challenges in practice, we designed and evaluated \textbf{ARC (Adaptive, Responsive, Closed-form Care Assistant)}, a voice-based AI care assistant for older adults---a population with distinct health and privacy considerations due to increased vulnerability and involvement of care partners. ARC is adaptive across interactions, responsive within interactions, and closed-form around predefined intervention goals and processes. ARC operationalizes multidomain lifestyle intervention as goal-directed conversational sessions targeting cognitive, mental, social, and physical domains of well-being. Its session protocols draw on  evidence-based approaches including cognitive strategy adaptation training\cite{bottiroli_importance_2013, bottiroli_self-guided_2017}, behavioral activation and motivational interviewing\cite{polenick_behavioral_2013, balan_integrating_2016}, and reminiscence therapy with life review\cite{huang_reminiscence_2015, lin_looking_2024}. We operationalize personalization as the extent to which ARC’s conversational behavior reflects domain- and participant-specific context and fidelity to evidence-based intervention delivery as whether ARC delivers the protocol-defined structure and evidence-based intervention components while avoiding off-protocol behavior. In a 3-day remote study, 80 older adults completed 473 sessions across six session types (9,183 total turns), alongside behavioral and self-report measures. Although ARC is informed by aging, MCI, and clinical intervention research, this formative study evaluates a broader older-adult population and does not assess clinical effectiveness or support generalization to older adults with MCI. This staged approach allows us to examine ARC's usability, acceptability, and process-level performance before evaluation with a more clinically vulnerable population.

Our analyses are guided by three research questions:
\begin{itemize}
    \item \textbf{RQ1 (User Level)}: How do older adults engage with and perceive ARC across repeated interactions?
    \item \textbf{RQ2 (System Level)}: To what extent does ARC conversational behavior reflect participant context while maintaining fidelity to evidence-based intervention delivery?
    \item \textbf{Exploratory RQ3 (Direct Care Ecosystem Level)}: How comfortable are older adults with different types of personal-data use and sharing across the AI assistant, clinicians, and care partners?
\end{itemize}

Older adults engaged in sustained, multi-turn interactions (19.4 turns, 6 minutes per session, on average) and reported favorable usability, acceptability, and intention to use, supporting the feasibility of this form of conversational intervention delivery. System-level analyses showed stronger domain-level than participant-specific contextualization, no detected systematic association between participant-specific contextual reuse and fidelity, and fidelity limitations were most evident in incomplete delivery of required session-structure and evidence-based components. Exploratory findings showed that expectations for personal-data use and sharing varied across data types and recipients, with participants generally reporting greater comfort with data used by ARC for personalized support than with sharing data with human care stakeholders.

This work makes three contributions:
 \begin{itemize}
     \item We operationalize and formatively evaluate multidomain lifestyle intervention for generative conversational delivery, translating intervention goals, participant context, and evidence-based components into multi-turn interactions.
     \item We provide process-level methods for evaluating contextual personalization and fidelity to evidence-based intervention delivery, including embedding-based contextual-reuse analyses that distinguish domain- from participant-specific contextualization and log-based fidelity analyses that characterize session-structure and evidence-based components coverage, and exploratory off-protocol drift.
     \item We characterize context-dependent expectations around data use and sharing across AI, clinicians, and care partners, highlighting design considerations for making data use legible and supporting context-sensitive control.
\end{itemize}

Together, this work informs the design, longitudinal deployment, and real-world evaluation of generative conversational agents for health interventions: beyond user experience and health outcomes, evaluation should establish whether the intended intervention processes are actually delivered.

\section{Related Work}

\subsection{Existing Voice-Based Agents for Goal-Directed and Auditable Care}
Voice-based health systems support activities ranging from open-ended conversation to clinical workflows and structured support. To situate ARC, we distinguish two dimensions: \textit{interaction mode}, describing how systems initiate or respond to interactions, and \textit{interaction structure}, describing how conversations are organized (Table~\ref{tab:voice_taxonomy}). These categories provide a descriptive comparison rather than an exhaustive classification.

\begin{table}[H]
\centering
\caption{Voice-based conversational systems classified by interaction mode and structure. Categories are descriptive and may overlap.}
\label{tab:voice_taxonomy}
\footnotesize
\setlength{\tabcolsep}{3pt}
\begin{tabular}{@{}p{0.13\textwidth}p{0.28\textwidth}p{0.53\textwidth}@{}}
\hline
\textbf{Dimension} & \textbf{Category} & \textbf{Definition} \\
\hline
Mode & Reactive & Responds to discrete user requests. \\
& Responsive & Sustains multi-turn dialogue, adapting to user input. \\
& System-initiated & Initiates contact through schedules or triggers. \\
& Passive (ambient) & Processes interactions without directing them. \\
\hline
Structure & Command-based & Completes discrete tasks without a session sequence. \\
& Open-ended & Allows dialogue without prescribed session stages. \\
& Guided & Follows predefined questions or task steps. \\
& Therapeutically guided (closed-form) & Uses predefined goals, required assessment or intervention components, and endpoints while allowing flexible responses. \\
\hline
\end{tabular}
\end{table}

Open-ended voice-based conversational agents such as ChatGPT Health and Ash support flexible dialogue \cite{openai_introducing_2026, slingshot_ai_slingshot_2025}, while systems such as Hippocratic AI, MyndYou, and Ellipsis Health support targeted clinical workflows \cite{mukherjee_polaris_2024, myndyou_case_2021, karlin_digital_2025}. Ambient systems such as Abridge and DAX Copilot support clinical documentation \cite{abridge_ai_inc_abridge_2026, voice_automated_dax_2026}, while ElliQ, MemPal, and Talk2Care support older adults through social engagement, memory, and care communication \cite{broadbent_elliq_2024, maniar_mempal_2025, yang_talk2care_2024}. Appendix Figure~\ref{fig:lit} illustrates representative systems across these dimensions.

Of particular interest are \textit{therapeutically guided, closed-form} interactions, which organize sessions around predefined goals, required assessment or intervention components, and endpoints without requiring scripted responses. This raises the question of how such processes can be delivered through generative dialogue while supporting individual adaptation and preserving fidelity to the intended intervention.

\subsection{Multidomain Lifestyle Intervention: Clinical Evidence and Digital Delivery}
Older adults often face interconnected challenges across cognitive, mental, social, and physical domains of well-being, yet most digital health technologies target a single behavior or health domain, such as social isolation, physical activity, or medication adherence\cite{zhang_conversational_2025}. Multidomain lifestyle interventions instead address these domains together, including cognitive engagement, emotional well-being, social participation, and physical activity. Large trials including FINGER\cite{kivipelto_finnish_2013}, US POINTER\cite{baker_study_2024}, and SMARRT\cite{yaffe_systematic_2019} have evaluated these approaches in older adults at risk for cognitive decline, with evidence syntheses suggesting cognitive and quality-of-life benefits\cite{salzman_associations_2022, castro_multi-domain_2023, roach_multidomain_2025}. Digital delivery through web-based platforms\cite{de_souto_barreto_web-based_2021}, telehealth programs\cite{rao_delivering_2025}, and mobile applications\cite{mattli_digital_2025} offers scalability but struggles with sustaining engagement and adherence, translating complex intervention protocols into usable interactions, and maintaining fidelity while adapting interventions to real-world contexts\cite{castro_multi-domain_2023, kebede_digital_2022, bott_face--face_2019}. 

More recently, Brill et al.\cite{brill_toward_2025} developed a rule-based conversational agent to deliver a just-in-time adaptive intervention (JITAI)-based multidomain lifestyle intervention for older adults with cognitive decline, and Holfelder et al.\cite{alexa_holfelder_implementation_2026} extended this approach with wearable sensing to provide sensor-informed, personalized lifestyle support for older adults with MCI. While these studies demonstrate the feasibility of conversational delivery, intervention selection and content remained governed by predefined rules. Generative AI may enable greater flexibility in how protocol-defined components respond to participant context, but whether this flexibility preserves required intervention structure and components remains unclear. ARC examines this question through goal-directed sessions and process-level evaluation of contextual personalization and fidelity of evidence-based intervention delivery.

\subsection{Personalization and Treatment Fidelity in Conversational Health Agents}

Older adults vary widely in health needs, cognitive status, technology literacy, and preferences, making personalization important for adapting digital health support over time\cite{kebede_digital_2022, riadi_digital_2022, graham_older_2021}. Established approaches like JITAIs tailor intervention timing or content to changing users' contexts and states using predefined decision rules\cite{nahum-shani_just--time_2018, klasnja_micro-randomized_2015}, while conversational health agents personalize feedback,  reminders, recommendations, and conversational style using user information and dialogue history\cite{kocaballi_personalization_2019}. Generative conversational agents, however, introduce a different measurement challenge: providing user context does not necessarily establish whether and how that context shaped the resulting interaction. Yet, evaluations of conversational health agents have largely focused on short-term interactions in controlled settings, feasibility, usability, and user experiences, rather than examining personalization in agents' conversational behavior\cite{xue_evaluation_2023}. 

Treatment fidelity, which is grounded in behavior and clinical intervention research, provides a complementary lens for evaluating what an intervention system actually delivers. The NIH Behavior Change Consortium identifies five components of treatment fidelity: study design, clinical provider training, treatment delivery, treatment receipt, and enactment\cite{bellg_enhancing_2004, borrelli_assessment_2011}. Our work focuses on treatment delivery—whether the intervention is delivered as intended. Related frameworks assess adherence to prescribed intervention content and processes\cite{carroll_conceptual_2007}, distinguish adherence from the competence or quality of delivery\cite{waltz_testing_1993, perepletchikova_treatment_2007}, and examine whether expected and proscribed therapeutic behaviors occur in clinicians communications during digital intervention delivery\cite{hadjistavropoulos_development_2018, zainal_capitalizing_2025}. 

Generative AI dialogue complicates intervention delivery fidelity because intervention processes unfold dynamically rather than through fixed content sequences; fidelity therefore cannot be inferred from design alone. The principle of \textit{flexibility within fidelity} calls for adapting delivery while preserving core therapeutic processes\cite{fonagy_fidelity_2019, zilcha-mano_evidence-based_2022}. Accordingly, we examine whether ARC's conversational behavior reflects participant context while maintaining predefined session structure and evidence-based components.

\subsection{Voice-Based Conversational Agents in Sociotechnical Care Ecosystems}
Personalized health support through generative voice-based conversational agents may require retaining and reusing information about users' histories, preferences, and needs across interactions. As these systems become embedded in older adults' \textit{direct care ecosystems}---that is, alongside their clinicians and care partners---the same information may be used by the AI for personalization or shared with clinicians and care partners for different care purposes. Yet users may have limited understanding of how voice-based systems collect, use, and share their information, even when they report general comfort with them\cite{cuadra_privacy_2026, lau_alexa_2018}. Prior work on "(un)usable privacy" further shows that technical approaches to privacy may not align with users' expectations in practice\cite{iravantchi_sok_2025, nagaraj_dissecting_2023-1}. Thus, it remains unclear how older adults evaluate information flows within their direct care ecosystem, limiting guidance for designing context-sensitive data use and sharing in AI-mediated care.

Contextual Integrity (CI)\cite{nissenbaum_privacy_2004, nissenbaum_contextual_2019} offers a useful lens for examining these context-dependent information flows by framing privacy in terms of whether information flows are appropriate given the actors, information types, and contexts involved. Rather than treating privacy preferences as uniform, we use ARC as a probe to examine older adults' comfort with personal-data use and sharing across different data types and recipients, including the AI assistant, clinicians, and care partners.

\section{Design of ARC for Multidomain Lifestyle Intervention}
\begin{figure}[tbp]
  \centering
  \includegraphics[width=\linewidth]{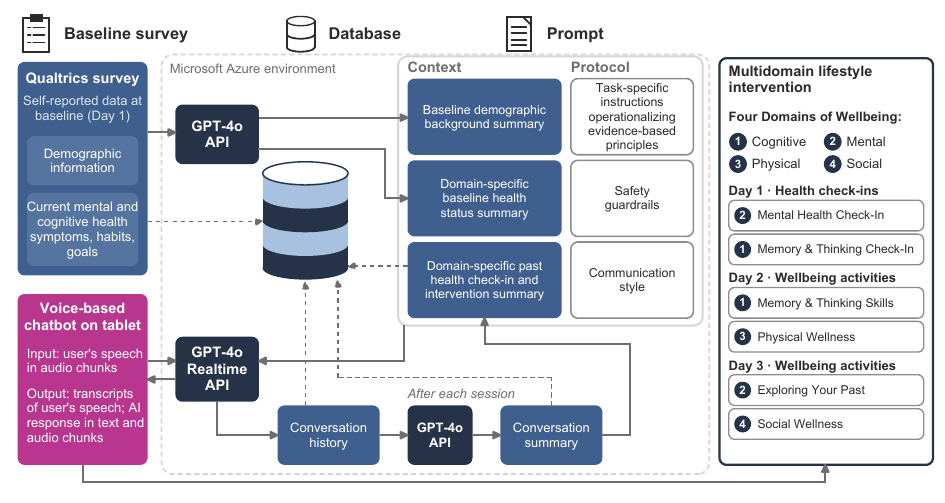}
  \caption{\textbf{ARC system architecture.} GPT-4o generates baseline profiles from Day~1 survey data. Profiles and prior-session summaries inform prompts that specify evidence-based processes, safety guardrails, and communication style. Participants converse by voice on a tablet using the GPT-4o Realtime API; summaries of each session inform later sessions. Numbers identify session well-being domains.}
  \Description{Left-to-right diagram from the baseline survey through GPT-4o profile generation and a database to the prompt (context and protocol), the realtime voice loop with the tablet chatbot, post-session summarization, and the six sessions over three days.}
  \label{fig:system_pic_and_study_flow}
\end{figure}

ARC (\textit{Adaptive, Responsive, Closed-form Care Assistant}) operationalizes multidomain lifestyle intervention for older adults through semi-structured generative conversations guided by predefined intervention goals, required processes, and endpoints. It incorporates prior participant context to adapt across sessions and responds flexibly to input within sessions (Figure~\ref{fig:system_pic_and_study_flow}). Research on aging, mild cognitive impairment (MCI), cognitive science, and clinical interventions informs \textit{what} ARC delivers across cognitive, mental, social, and physical domains, while participant context informs \textit{how} these processes are delivered.

\subsection{Speech-to-Speech Conversational Agent Design}
\label{speech-to-speech-ca-design}
The system was implemented with a JavaScript frontend and a Python FastAPI backend, deployed within a secure Microsoft Azure environment (Figure~\ref{fig:system_pic_and_study_flow}). We used Azure OpenAI GPT-4o-Realtime\cite{microsoft_azure_azure_2024} (2024-12-17) to support low-latency, bidirectional speech interaction with automatic transcription on tablet devices. Turn-taking was handled via server-side voice activity detection (VAD; threshold = 0.7, 300 ms prefix padding, 3 s silence window), tuned to accommodate older adult speech patterns (e.g., slower pacing, longer pauses, softer onsets). All processing occurred within the Azure environment, with audio handled in memory and not stored at rest. Backend services supported system prompting, integration of participants’ baseline profiles and prior sessions, and logging of interactions across study sessions. At the end of each session, we generated (1) a full transcript and (2) a structured summary using a separate GPT-4o model\cite{openai_gpt-4o_2024} (2024-11-20), capturing high-level features (e.g., session overview, conversational valence, safety signals). These outputs supported monitoring of system behavior and participant safety. All data were transmitted over HTTPS and stored securely with access restricted to authorized study personnel.

\subsection{Contextual Personalization}
ARC incorporated two sources of participant context into system prompts: baseline profiles and summaries of prior domain-relevant sessions. These inputs were intended to support continuity and adaptation across interactions; RQ2 evaluates whether they were reflected in ARC's responses. Baseline profiles summarized domain-specific health status, goals, and self-management strategies (Figure~\ref{fig:system_pic_and_study_flow}). Demographic attributes were included in prompts but excluded from personalization analyses because they were unlikely to be reflected in conversational content.

\subsubsection{Baseline Profile Generation}
Baseline profiles were generated from a Day-1 Qualtrics survey capturing demographic information and domain-specific health status across mental, cognitive, social, and physical well-being. Backend services processed survey responses at the first chatbot session, scoring validated questionnaires and structuring outputs into profiles provided as contextual input for subsequent interactions (Figure~\ref{fig:system_pic_and_study_flow}) using a GPT-4o model (2024-11-20)\cite{openai_gpt-4o_2024}.

Mental health profiles included Geriatric Anxiety Scale–10 (GAS-10) and Geriatric Depression Scale–15 (GDS-15) scores and endorsed symptoms\cite{mueller_geriatric_2015, yesavage_9geriatric_1986, brown_development_2005}. Cognitive health profiles were derived from the Everyday Cognition Scale (ECog-41), including subscale scores across memory, language, visuospatial abilities, planning, organization, and attention\cite{farias_revisions_2020}. Physical and social profiles captured self-reported medical conditions, functional limitations, physical activity and sleep, social engagement (e.g., visiting friends, volunteering), transportation use, and current health management strategies and goals. As wearable sensing was not collected, participants estimated daily step count and sleep duration using guided ranges (e.g., $\sim$100 steps/minute). Together, these measures provided a structured snapshot of participants’ health status and routines that ARC could draw on during interaction.

\subsubsection{Prior Session Context Generation}
After each session, a GPT-4o model (2024-11-20)\cite{openai_gpt-4o_2024} generated a structured summary including session overview, health goals, action items, distress ratings, and safety assessments. Relevant fields from the most recent summaries were retrieved from cloud storage and incorporated into subsequent prompts to provide continuity across sessions. For health check-ins, only session overview and health goals were reused within the same check-in type (i.e., Mental Health Check-In summaries informed future Mental Health Check-Ins; Memory and Thinking Check-In summaries informed future Memory and Thinking Check-Ins). For well-being activities (i.e., lifestyle intervention sessions), these fields were also incorporated across domains: Mental Health Check-In summaries informed Reminiscence Therapy, Social Wellness, and Physical Wellness sessions, while Memory and Thinking Check-In summaries informed Memory and Thinking Skills Training sessions. Each intervention session also referenced its own prior summaries, including action items, when available.

\subsection{Intervention Protocol and Prompt Design}
The assistant was implemented as a modular prompt system with shared policies and session-specific procedures. Each session protocol specified its goal, required session-structure and evidence-based components, behavioral boundaries, and endpoint. Within this structure, the generative model could respond flexibly to participant input and context, allowing conversational adaptation while preserving the processes required by the session protocol. This design reflects the principle of \textit{flexibility within fidelity}: ARC could vary how intervention processes were delivered conversationally while maintaining the core processes each session was designed to deliver.

The system followed an older-adult–oriented communication style (short turns, simple language, warm tone, patient pacing), with explicit role boundaries (e.g., “AI tool, not a human clinician/caregiver”) and safeguards for potentially harmful inputs (e.g., self-harm, violence, medical advice, privacy concerns), emphasizing non-escalation, clear limitations, and redirection to session goals. We defined two session types with distinct prompting constraints: \textbf{(1) health check-ins and (2) multidomain lifestyle intervention sessions.}

\subsubsection{Health Check-Ins}
Health check-ins were assessment-oriented sessions targeting: \textbf{(1) Mental Health (mood and anxiety) and (2) Memory and Thinking (daily cognitive functions) check-ins}. Prompts constrained assessment to “the past few days” and instructed the assistant to ask about symptoms and everyday experiences in the context of baseline and prior-session summaries using relevant items from standardized instruments (e.g., GDS-15\cite{yesavage_9geriatric_1986, brown_development_2005}, GAS-10\cite{mueller_geriatric_2015}, and ECog-12\cite{manjavong_performance_2024}). Follow-up questions were limited to \textit{frequency, duration, intensity, timing, and functional impact}. To maintain engagement, the assistant used supportive clinical interviewing and Motivational Interviewing (MI) principles (e.g., reflective statements, open-ended questions about user experiences and goals) while remaining non-directive (e.g., not introducing goals or prescribing actions). Behaviors beyond this intended session-delivery process were considered shifts from assessment into unsolicited intervention.

\subsubsection{Multidomain Lifestyle Intervention Sessions}
The assistant delivered \textit{four semi-structured, AI-guided lifestyle well-being activities} targeting mental, cognitive, physical, and social well-being. The activities drew on intervention approaches used in aging, mild cognitive impairment (MCI), cognitive science, and clinical intervention research, including strategy adaptation training\cite{bottiroli_importance_2013, bottiroli_self-guided_2017}, behavioral activation (BA\cite{cuijpers_behavioral_2007}), motivational interviewing (MI\cite{balan_integrating_2016}), and reminiscence therapy with life review\cite{cammisuli_effects_2022, huang_reminiscence_2015, lin_looking_2024}. These approaches informed the evidence-based components required within each session type and later examined in our analysis of \textit{fidelity of intervention delivery} (see Appendix~\ref{evidence-based_intervention_principles} for detailed mappings). Sessions began with goal orientation and incorporated relevant participant context when available. 
\begin{enumerate}

\item \textbf{"Memory and Thinking Skills" (Strategy Adaptation Training + BA + MI)} 
This cognitive wellness activity was designed to support everyday functioning by helping participants learn memory strategies, recognize task demands, and adapt strategies to real-world tasks. Grounded in strategy-adaptation training\cite{bottiroli_importance_2013, bottiroli_self-guided_2017, hertzog_metacognitive_2012}, sessions introduced empirically supported techniques (e.g., interactive imagery, sentence generation) and guided \textit{task analysis} through a \textit{pause-and-plan} approach to scaffold metacognitive awareness. Participants practiced with personally relevant tasks (e.g., grocery lists, daily activities) to support transfer to everyday functioning.

\item \textbf{"Exploring Your Past" (Reminiscence Therapy with Life Review + BA + MI)} 
This activity was designed to support emotional well-being and memory through guided autobiographical reflection informed by reminiscence and life-review approaches\cite{cammisuli_effects_2022, lin_looking_2024}. Sessions followed a three-phase structure (\textit{recall}, \textit{reflection}, optional \textit{integration}), using open-ended prompts to elicit participant-led storytelling and meaning-making. BA framed reminiscence as meaningful engagement, while MI supported autonomy and reflection. When appropriate, the assistant facilitated gentle integration of past experiences with present identity without imposing interpretation.

\item \textbf{"Social Wellness" (BA + MI)} 
This activity aimed to strengthen social connections using a combined BA+MI framework\cite{balan_integrating_2016, pepin_modifying_2021}. Sessions included brief psychoeducation, reflection on current social routines, and identification of meaningful social domains (e.g., family, community). The assistant supported goal-setting by co-creating small, feasible, values-based social goals and translating them into concrete implementation intentions.

\item \textbf{"Physical Wellness" (BA + MI)} 
This activity applied the same BA+MI framework to physical health behaviors (e.g., physical activity, sleep, stress)\cite{friederichs_motivational_2015, lilienthal_telephone-based_2014, tse_motivational_2013}. Sessions followed the same structure—goal orientation, reflection, and action planning—tailored to participants’ physical capabilities and preferences.

\end{enumerate}

\subsection{Safety Guardrails}
We implemented two safety layers: Azure OpenAI content filtering screened inputs and outputs for harmful content and blocked high-risk cases\cite{patrickfarley_content_2026}, while application-level rule-based protocols addressed self-harm, medical advice, misinformation, abuse, and privacy concerns. These protocols instructed ARC to avoid amplifying harmful content, state limitations, acknowledge emotions without endorsing harmful intent, and redirect to appropriate support or session goals. Full specifications are provided in Appendix~\ref{appendix:safety_protocols}.

\section{Methods}\label{sec:Methods}

\subsection{Participants} \label{sec:Methods:Ptt}
The study was approved by the authors’ institutional review board. Participants were recruited via Prolific and were eligible if they were aged 65 or older, fluent in English, and able to complete the study on a tablet device. Because ARC's longer-term target population includes older adults with MCI, we intentionally first evaluated the interaction paradigm in a broader sample of adults aged 65 and older. This staged evaluation was designed to identify system- and interaction-level limitations before introducing additional usability, comprehension, and safety considerations associated with cognitive impairment. Accordingly, the present sample is not treated as representative of older adults with MCI. All participants provided informed consent and completed up to six interaction sessions over three study days (approximately 30 minutes each), receiving \$6 for Day 1, \$6 for Day 2, and \$8 for the optional Day 3.

Of 98 enrolled participants, four did not complete the Day 1 chatbot session, three withdrew on Day 2, two did not complete Day 2, eight did not complete Day 3, and one was excluded due to persistent audio issues. Analyses include participants who completed the study procedures across all three days ($N=80$). Participants were community-dwelling older adults (ages 65–80, $M=70.49$, $SD=3.85$), predominantly female (67.5\%) and White (87.5\%), with diverse education, income, and employment backgrounds; most were married (56.2\%) or retired (51.2\%) (Table~\ref{tab:table1_demographics}).

To characterize baseline mental and cognitive status, participants completed the Geriatric Anxiety Scale-10 (GAS-10 \cite{mueller_geriatric_2015}), Geriatric Depression Scale-15 (GDS-15 \cite{yesavage_9geriatric_1986, brown_development_2005}), and Everyday Cognition Scale (ECog-41; revised self-report \cite{farias_revisions_2020}), assessing anxiety, depression, and everyday cognitive functioning (e.g., memory, language, planning, attention). Participants reported generally low-to-moderate anxiety and depressive symptoms and relatively preserved everyday cognitive functioning. They also reported high mobile-device proficiency (MDPQ-16 \cite{roque_new_2018}) and frequent technology use, including smartphones (98.8\%). Most had prior experience with LLM-based conversational agents (88.8\%) and voice assistants (66.2\%). Detailed baseline health, cognition, and technology-familiarity measures are reported in Appendix~\ref{app:participant_characteristics}.

\subsection{Study Design and Procedures}\label{sec:Study Design}
This formative, single-arm deployment-and-evaluation study examined the feasibility and interactional processes of ARC, a voice-based AI assistant delivering a multidomain lifestyle intervention. We evaluated (1) older adults' engagement, usability, acceptability, and intention to use; (2) contextual personalization and fidelity of evidence-based intervention delivery; and (3) exploratory expectations for personal-data use and sharing across ARC, clinicians, and care partners. The single-arm design supported evaluation of system behavior and intervention delivery rather than comparative assessment of clinical effectiveness; therapeutic and health outcomes were not measured.

\begin{figure}[tbp]
  \centering
  \includegraphics[width=0.82\linewidth]{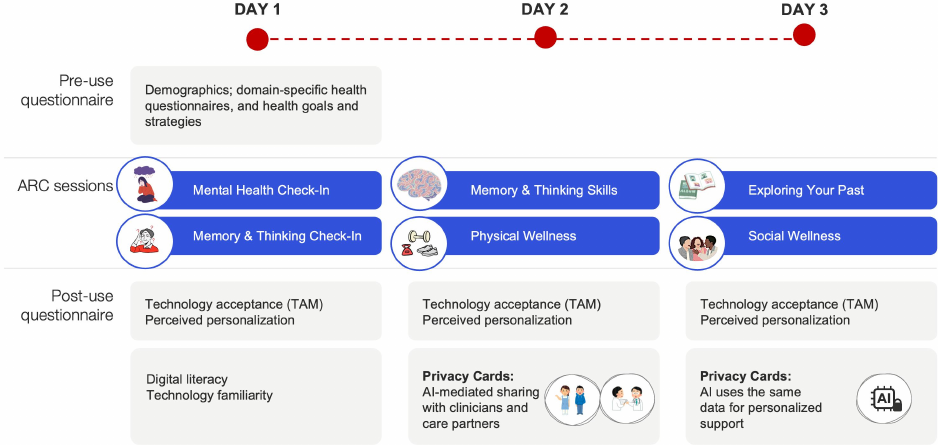}
  \caption{\textbf{Study procedure.} Participants completed a pre-use survey on Day~1, two ARC sessions per day for three days, and post-use surveys each day. Privacy Cards were administered on Days~2 and~3.}
  \Description{Three-column timeline for Days 1 to 3 with rows for the pre-use questionnaire, the ARC sessions, and the post-use questionnaires.}
  \label{fig:StudyDesign}
\end{figure}

As shown in Figure~\ref{fig:StudyDesign}, the study spanned three days, with Qualtrics surveys and two ARC sessions per day: pre-chatbot surveys assessing demographic and health status, mental and cognitive health check-ins sessions (Day 1), physical wellness and memory and thinking skills (Day 2), and social wellness and reminiscence-based activities (Day 3). Post-chatbot surveys assessed technology acceptance, perceived personalization, and, on Days 2-3, comfort with personal-data use and sharing using Privacy Card scenarios (Figure~\ref{fig:privacy_cards}; see further details in Section \ref{privacy_cards_measure}). Before their first session, participants were informed that ARC was \textit{not a human or medical professional} and that only the research team could access their data; comprehension checks confirmed understanding.

\subsection{Measures and Materials}\label{sec:Measures}
\subsubsection{RQ1: Engagement, Perceived Acceptability, Usability, and Intention to Use}
Behavioral engagement was assessed using conversational turn counts and session duration, excluding one session with invalid timestamps from duration analyses. Technology acceptance was measured using the Technology Acceptance Model (TAM \cite{holden_technology_2010}), comprising perceived usefulness (8 items), ease of use (3 items), attitude toward the technology (3 items), and behavioral intention to use (6 items). All items used a 5-point Likert scale (1 = Strongly Disagree, 5 = Strongly Agree). All constructs demonstrated high internal consistency across study days (Cronbach's $\alpha=0.89$-$0.98$; Appendix~\ref{app:rq1}).

\subsubsection{RQ2a: Personalization}
We examined two complementary aspects of personalization: observable contextual reuse and participants' perceived personalization. Contextual reuse examined how ARC's responses reflected domain- and participant-specific information from baseline profiles and prior-session summaries (see Section~\ref{rq2a_personalization_analysis}). Perceived personalization was assessed after each study day using three items measuring perceived tailoring to health status, preferences, and abilities, rated on a 5-point Likert scale (1 = Strongly Disagree, 5 = Strongly Agree).

\subsubsection{RQ2b: Fidelity of Intervention Delivery}
Drawing on established treatment-fidelity literature, we focus on one dimension of fidelity relevant to ARC: \textit{fidelity of intervention delivery}, or whether the intervention is delivered as intended \cite{bellg_enhancing_2004, borrelli_assessment_2011, carroll_conceptual_2007}. 
We do not assess broader fidelity dimensions such as treatment receipt and enactment, or the competence or quality with which intervention components are delivered. 
We operationalize intervention-delivery fidelity directly from conversational logs through two primary coverage measures and a secondary exploratory measure of protocol drift. \textit{Session-structure coverage} assessed the proportion of required structural components delivered, while \textit{evidence-based component coverage} assessed the proportion of required evidence-based components delivered. \textit{Drift rate} captured the proportion of assistant turns that departed from the prescribed session protocol. For assessment-oriented check-ins, evidence-based components included assessment content and practices derived from validated measures; for intervention sessions, they included components derived from the session's corresponding evidence-based intervention approaches.
Section \ref{rq2b_fidelity_analysis} describes the annotation, validation, and statistical procedures used to derive these measures.

\subsubsection{Exploratory RQ3: Privacy Cards for Contextual Privacy Judgment}
\label{privacy_cards_measure}
\begin{figure}[tbp]
  \centering
  \includegraphics[width=\linewidth]{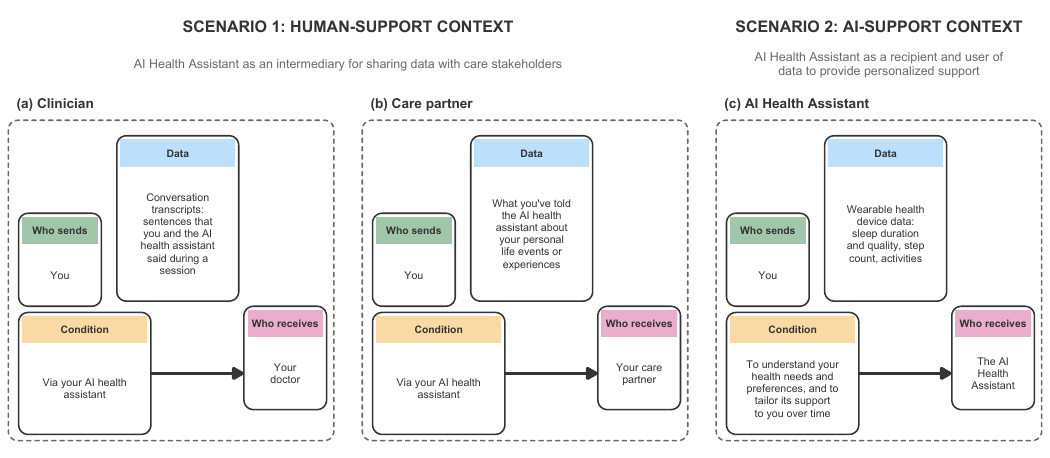}
  \caption{\textbf{Privacy Cards for exploratory RQ3.} Cards specify the data source, data type, and proposed use or recipient. Scenario~1 (human support) describes sharing through ARC with (a)~a clinician or (b)~a care partner; Scenario~2 (AI support) describes (c)~use by ARC for personalization.}
  \Description{Three example cards: data sent via the AI health assistant to a doctor, to a care partner, and used by the AI health assistant itself.}
  \label{fig:privacy_cards}
\end{figure}

We used a modified \textit{Privacy Cards} method \cite{cuadra_privacy_2026} to examine context-dependent perceptions of appropriate data sharing and use across seven data types and three recipients: clinical providers (e.g., psychologist, primary care physician), care partners (e.g., family member, home aide), and the \textit{AI care assistant} (Figure~\ref{fig:privacy_cards}). Data types included demographic information, chat history, AI-generated summaries, time-based activity data, voice/audio recordings, wearable-derived data, and reminiscence therapy content. Privacy Cards make information flows explicit by specifying data type, recipient, and purpose. While informed by Contextual Integrity (CI) \cite{nissenbaum_privacy_2004, nissenbaum_contextual_2019}, our use of the method captures participants' subjective judgments of appropriateness rather than eliciting formal privacy norms around when AI systems mediate or enact personalized care processes. 

For each data type, participants evaluated two contexts. In the \textit{human-support context}, they rated comfort with data shared \textit{via} the AI care assistant to clinicians and care partners (AI as intermediary). In the \textit{AI-support context}, they rated comfort with data \textit{used by} the AI care assistant to personalize interaction (AI as active care participant). Comfort was rated on a 7-point Likert scale (1 = \textit{Very Uncomfortable}, 7 = \textit{Very Comfortable}). By varying data type and recipient while holding purpose constant (personalized support), this method captures \textit{context-dependent judgments of appropriateness}.

\subsection{Data Preprocessing and Analysis}\label{sec:Data Preprocessing and Analysis}
Across three days, 80 participants completed six session types. A deployment error produced 67 duplicate files for 39 participants (48\%). We retained the earliest valid completion per participant and session type, defined as at least eight turns, using file timestamps; session types without a valid completion were treated as missing. This yielded 473 retained sessions and 9,183 conversational turns. 

\subsubsection{RQ1: Engagement, Perceived Acceptability, Usability, and Intention to Use}\label{rq1_analysis}
Given the formative, single-arm design, RQ1 analyses were descriptive rather than causal. We examined turn counts and session duration by session type and day using repeated-measures ANOVA. TAM constructs (usefulness, ease of use, attitude, and behavioral intention) were summarized descriptively and compared across days using repeated-measures ANOVA with corrected post-hoc tests where applicable; Pearson correlations examined associations among constructs.

\subsubsection{RQ2a: Contextual Reuse Analysis and Perceived Personalization}\label{rq2a_personalization_analysis}
To evaluate whether the AI’s responses reflected personalized context, we used a sentence embedding approach with a three-condition comparison design, testing (1) domain-specificity and (2) participant-specificity. 

\paragraph{Personalization Context Construction}
For each session, the personalization context consisted of domain- and participant-specific information provided via system prompts (Appendix, Figure~\ref{fig:personalization_contexts}), including baseline profiles and, where applicable, summaries of prior related sessions. Contexts were mapped to session types by target domain (mental, cognitive, physical, social; Appendix~\ref{personalization_contexts_appendix}). Demographic information was excluded as it is unlikely to be reflected in conversational content.

\paragraph{Embedding and Similarity} 
Context strings and assistant turns were encoded using the \textit{all-mpnet-base-v2} sentence-transformer model \cite{reimers_sentence-bert_2019}. Embeddings were L2-normalized, making cosine similarity equivalent to the dot product. Similarity was computed at the turn level and averaged across assistant turns to produce a session-level score per condition, treating the session as the unit of analysis. 

\paragraph{Comparison Design} 
We compared three context conditions (Figure~\ref{fig:personalization_contexts}). The \textit{within-domain, within-participant} condition (target) uses the participant's context for the session's target domain. Two controls were used: (1) \textit{cross-domain, within-participant}, using the same participant's contexts for all three non-target domains, each built the same way as the target, to test domain-specificity (we report the mean across the three as the primary control and the closest non-target domain as a stricter sensitivity check), and (2) \textit{within-domain, cross-participant}, using leave-one-out mean similarity to other participants’ corresponding domain contexts to test participant-specificity. Together, these comparisons isolate whether the system uses the correct domain context and whether it reflects participant-specific information beyond general patterns. The cross-participant condition uses other participants’ injected context strings (baseline profiles and check-in summaries), not their conversation turns.

\paragraph{Statistical Analysis} 
Session-level mean cosine similarities were compared using two-tailed Wilcoxon signed-rank tests: within-domain, within-participant versus cross-domain, within-participant (domain specificity), and versus within-domain, cross-participant (participant specificity). We report Cohen's $d_{\mathrm{av}}$ (mean paired difference divided by the mean condition standard deviations). Session-type analyses used Benjamini-Hochberg FDR correction within each six-test comparison family, reporting raw $p$ and adjusted $q$ values. Pooled analyses were primary and were not corrected across session types.

\paragraph{Perceived Personalization}
For perceived personalization, we used a $3$ (Dimension: health status, preferences, abilities) $\times$ $3$ (Day) repeated-measures ANOVA, with Greenhouse-Geisser correction where applicable and Bonferroni-corrected post-hoc comparisons.

\subsubsection{RQ2b: Fidelity of Intervention Delivery}
\label{rq2b_fidelity_analysis}
A generative AI-based conversational assistant may follow the overall session structure while still omitting evidence-based components (e.g., motivational interviewing processes such as evocation or planning). Conversely, it may deliver relevant evidence-based intervention content while departing from the prescribed session process. Drawing on treatment-fidelity research distinguishing prescribed and proscribed therapeutic behaviors\cite{waltz_testing_1993, zainal_capitalizing_2025}, we therefore examined two complementary aspects of intervention delivery: (1) coverage of required session components and (2) departures from the intended session process.

\paragraph{Annotation Scheme and Codebook Development}
We developed a multi-label taxonomy for characterizing how assistant turns followed or departed from each session protocol. Because individual turns could serve multiple functions, multiple labels could be assigned to the same turn.
\begin{itemize}
    \item \textit{On-task}: protocol-aligned turns that advanced the session goal, with subtypes including assessment, reflection, planning, and training.
    \item \textit{Drift}: turns that departed from the session goal or protocol, including \textit{content} drift (off-goal or non-prescribed content), \textit{functional} drift (a disallowed action for that session), \textit{process} drift (procedural deviations such as bundled unrelated questions or premature closure), \textit{boundary} violations, and \textit{misgrounded} responses.
    \item \textit{Recovery}: return toward the intended process after drift, coded as \textit{explicit}, \textit{implicit}, or \textit{partial}. Recovery was included in the annotation scheme but was not examined quantitatively because automated recovery labels were too sparse to support reliable estimates.
\end{itemize}

To develop the coding manual, three annotators (the first author and two co-authors) independently coded 60 sessions---10 randomly selected sessions from each of the six session types, comprising 578 assistant turns. The annotators reviewed disagreements to clarify ambiguous distinctions and iteratively refine the coding definitions. Using the finalized coding manual and the three sets of annotations, the first author, who has prior clinical and behavioral health background, adjudicated the final reference labels for these sessions.

\paragraph{LLM-Assisted Annotation}
Because manual coding of the full conversation corpus was not feasible, we used GPT-5.4-mini to apply the finalized coding scheme using few-shot in-context learning. For each session type, four reference-labeled sessions were provided as worked examples and six were held out for validation, yielding 24 in-context examples and 36 held-out validation sessions (352 assistant turns). The model received the finalized coding manual, corresponding session protocol, and four examples from the same session type. On the held-out validation set, model-generated labels aligned with the human reference labels at micro-averaged Cohen's $\kappa=0.70$. The model generated both turn-level drift labels and separate episode-level records identifying the start and end of deviations. These outputs were not consistently aligned. We therefore used turn-level labels for exploratory drift-rate analyses rather than the separately generated episode records.

\paragraph{Independent Human Audit}
As an additional validation step, one co-author who was not involved in reference-label development independently annotated 12 previously unseen conversations (two from each session type) using the finalized coding manual while blinded to the LLM-generated labels. Because identifying deviations from evidence-based assessment and intervention processes requires domain-specific knowledge, disagreements were subsequently reviewed with the first author, who has prior training and experience in psychodiagnostic and neurocognitive assessments and in behavioral health intervention research. Adjudication was blinded to LLM labels and conducted against the predefined coding criteria and session protocols to establish final human reference labels. We compared the LLM-generated annotations with these adjudicated labels. Agreement was good for session-structure coverage (ICC $=0.82$, 95\% CI [0.49, 0.95]) and moderate-to-good for evidence-based component coverage (ICC $=0.74$, 95\% CI [0.34, 0.91]). Drift-label agreement was assessed separately against the adjudicated human reference labels. Because drift annotations were less reliable than the coverage measures, drift-based analyses were treated as exploratory.

\paragraph{Fidelity of Intervention Delivery Metrics}
We defined three complementary session-level metrics. The two primary measures were \textit{session-structure coverage}, the proportion of required structural components delivered, and \textit{evidence-based component coverage}, the proportion of required components derived from the relevant assessment or intervention approaches that were delivered. We additionally examined \textit{drift rate}, the proportion of assistant turns labeled as departing from the session protocol, as a secondary exploratory measure.

\paragraph{Statistical Analysis across Session Types}
Fidelity analyses used the retained conversation corpus. Of the 473 retained sessions, 23 were excluded because they had served as in-context examples for the automated annotator and therefore had no automated labels. The remaining in-context example came from a participant whose data were not included in the retained corpus. One additional session was excluded due to a technical glitch, leaving 449 sessions from 80 participants for analysis. The same sessions were used in the personalization-fidelity models below.

To examine differences across session types while accounting for repeated sessions from the same participant, we fit binomial-logit mixed-effects models with session type as a fixed effect and participant as a random intercept. Session-structure and evidence-based component coverage were modeled as components present out of components required. Drift rate was modeled as turns labeled as drift out of total assistant turns. Because drift counts were overdispersed relative to the binomial distribution, the drift model additionally included an observation-level random effect. To characterize when drift occurred, we normalized each assistant turn's position within its session (0\% = first, 100\% = last assistant turn) and grouped positions into ten equal bins. For each session and bin, we computed the share of assistant turns labeled as drift, then averaged these session-level rates within each session type; 95\% confidence intervals were obtained from 2,000 participant-level bootstrap resamples. We also summarized drift episodes, defined as runs of consecutive drift-labeled assistant turns, and the drift subtypes they contained. These analyses are descriptive; differences across conversational progress were not formally tested.

\paragraph{Relationship Between Personalization and Fidelity}
In a follow-up exploratory analysis, we examined whether participant-specific contextual reuse was associated with fidelity of intervention delivery. We focused on session-structure coverage and evidence-based component coverage as the two primary outcomes, with drift rate included as a secondary exploratory measure. For each outcome, we fit a separate mixed-effects model with participant-specific contextual reuse and session type as fixed effects and participant as a random intercept. The drift model additionally included an observation-level random effect. Sensitivity analyses adjusted for conversation length.

\subsubsection{Exploratory RQ3: Context-Dependent Perceptions of Appropriate Data Sharing and Use}
To examine context-dependent perceptions of appropriate data use and sharing, participants rated their comfort of sharing seven data types across three recipients: clinicians, care partners, and the AI care assistant. We tested whether comfort varied by recipient and data type using a $3$ (Recipient) $\times$ $7$ (Data Type) repeated-measures ANOVA. Greenhouse-Geisser corrections were applied where appropriate, generalized eta squared ($\eta^2_G$) was reported as the effect size, and post-hoc pairwise comparisons were Bonferroni-corrected. To compare the two data-use contexts represented in the Privacy Cards, we conducted a planned paired-samples $t$-test comparing participants' mean comfort with \textit{human-support sharing} (data shared via the AI with clinicians and care partners, averaged across recipients and data types) and \textit{AI-support use} (data used by the AI for personalization, averaged across data types).

\section{Results}
\subsection{RQ1: Engagement, Perceived Acceptability, Usability, and Intention to Use}
\subsubsection{Older Adults Sustained Multi-Turn Interactions Across Session Types}

Across all retained sessions ($n=473$), interactions averaged $19.4$ turns ($SD=7.0$), or $6.0$ minutes ($SD=2.4$) excluding one Social Wellness session with invalid timestamps. Interaction length varied across session types, with check-in sessions generally shorter than intervention sessions. Memory and Thinking Skills involved the most turns ($M=25.0$) and was among the longest in duration ($M=7.2$ minutes), while Exploring Your Past sessions were similarly long ($M=7.1$ minutes). Full descriptive statistics by session type are reported in Appendix \ref{app:rq1}.

\subsubsection{Older Adults Reported Generally Positive Perceptions of Usefulness, Ease of Use, and Intentions to Use}
\begin{figure}[tbp]
  \centering
  \includegraphics[width=\linewidth]{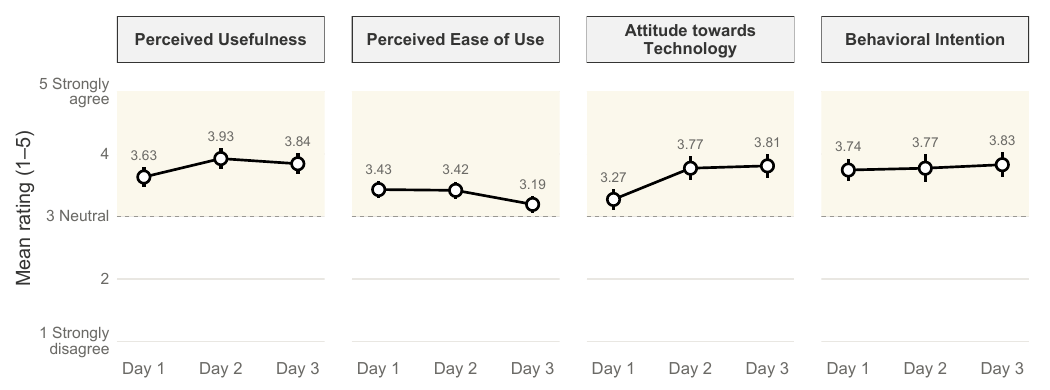}
  \caption{\textbf{Technology acceptance by study day ($N=80$).} Mean ratings (1--5) with 95\% confidence intervals. Shading indicates scores above the neutral midpoint.}
  \Description{Four small line charts of mean usefulness, ease of use, attitude, and intention ratings on Days 1 to 3, all above the midpoint.}
  \label{fig:tech_acceptance}
\end{figure}

Technology acceptance was generally favorable across study days (Figure~\ref{fig:tech_acceptance}). Mean ratings exceeded the scale midpoint for usefulness ($M=3.63$-$3.93$), ease of use ($M=3.19$-$3.43$), attitudes ($M=3.27$-$3.81$), and intention to use ($M=3.74$-$3.83$). Usefulness and attitudes improved modestly after Day 1, while intention to use remained stable. TAM constructs were positively correlated (Appendix~\ref{app:rq1}).

\subsection{RQ2a: Contextual Personalization}
\subsubsection{ARC Showed Strong Domain-Level but Modest Participant-Specific Contextualization}
\begin{figure}[tbp]
  \centering
  \includegraphics[width=\linewidth]{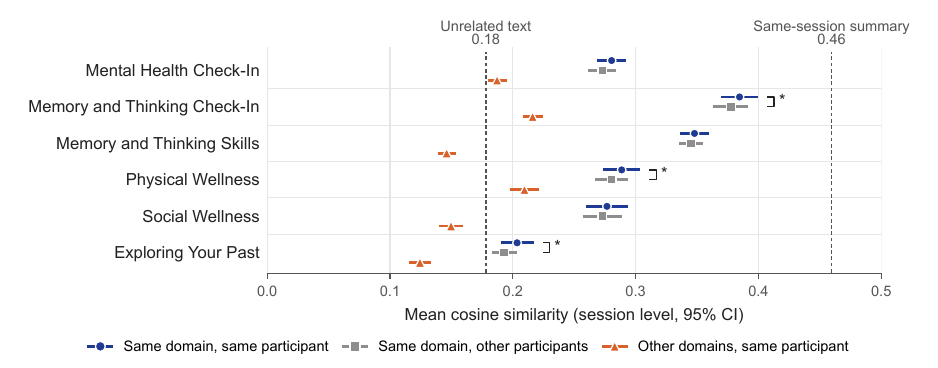}
  \caption{\textbf{Contextual reuse by session type (473 sessions).} Mean session-level cosine similarity (95\% CI) between ARC responses and three contexts: same participant and domain, same participant across domains, and different participants within the domain. Dashed lines show reference similarity scores for unrelated text (lower) and same-session summaries (upper). $^{*}$Same participant $>$ other participants after FDR correction ($q<0.05$).}
  \Description{Dot-and-interval plot of similarity to the same-domain own context, other participants' same-domain context, and the participant's other-domain contexts for six session types.}
  \label{fig:rq2a-contextual-reuse}
\end{figure}

Across all retained sessions ($n=473$), ARC's responses showed substantially greater similarity to within-domain, within-participant context than to the same participant's context from the three other domains ($W=440$, $p<0.001$, $d=1.81$; closest other domain: $W=12{,}488$, $p<0.001$, $d=0.88$), indicating strong domain-level contextualization (Figure~\ref{fig:rq2a-contextual-reuse}). Responses were also more similar to the participant's own context than to same-domain context from other participants ($W=42{,}895$, $p<0.001$), but this participant-specific effect was small ($d=0.08$).

Domain-level contextualization was consistent across all six session types (all FDR-adjusted $q<0.001$; $d=1.32$-$4.78$). Against the closest other domain, the difference held in five of six session types ($d=0.35$-$3.37$) but not for Physical Wellness ($p=0.13$, $d=0.16$). Participant-specific contextualization was less consistent: after FDR correction, reliable effects remained only for Memory and Thinking Check-In, Physical Wellness, and Exploring Your Past, with small effect sizes ($d=0.11$-$0.20$). Full session-type comparisons are reported in Appendix \ref{personalization_contexts_appendix}.

\subsubsection{Older Adults Perceived ARC as Personalized Despite Modest Participant-Specific Contextual Reuse}
\begin{figure}[tbp]
  \centering
  \includegraphics[width=0.8\linewidth]{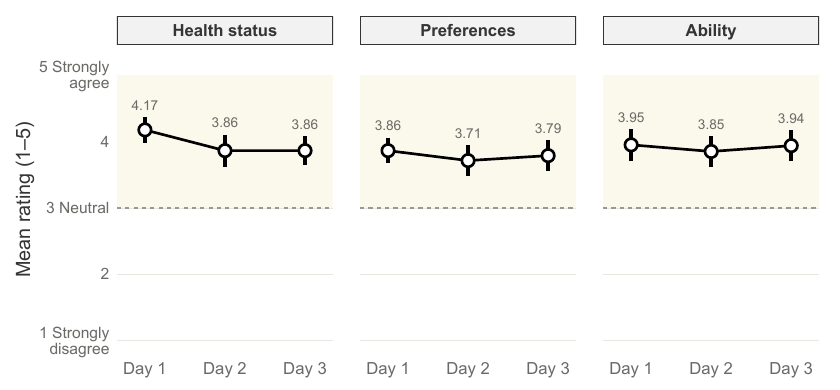}
  \caption{\textbf{Perceived personalization by study day ($N=80$).} Mean agreement (1--5; 95\% CI) that ARC tailored conversations to participants' health status, preferences, and ability.}
  \Description{Three small line charts of mean personalization ratings on Days 1 to 3, all above the midpoint.}
  \label{fig:perceived_personalization}
\end{figure}

Participants nevertheless reported generally positive perceptions of personalization across health status, preferences, and abilities (all means $>3$ on a 5-point scale; Figure~\ref{fig:perceived_personalization}). Perceived personalization differed modestly across dimensions, $F(2,158)=4.56$, $p=0.012$, $\eta^2_G=0.006$, with ratings highest for health status ($M=3.97$), followed by ability ($M=3.91$) and preferences ($M=3.79$). Health-status personalization was rated higher than preference-based personalization ($p=0.002$, Hedges' $g=0.23$). Perceptions did not vary significantly by study day or by the dimension $\times$ day interaction (both $p>0.10$).

Thus, participants consistently experienced ARC as personalized even though participant-specific contextual reuse in the conversational logs was comparatively modest. These findings suggest that perceived personalization and observable participant-specific contextualization capture related but distinct aspects of personalization.

\subsection{RQ2b: Fidelity of Evidence-Based Intervention Delivery}
\subsubsection{Fidelity Limitations Were Most Apparent in Incomplete Delivery of Required Components}
\begin{figure}[tbp]
  \centering
  \includegraphics[width=\linewidth]{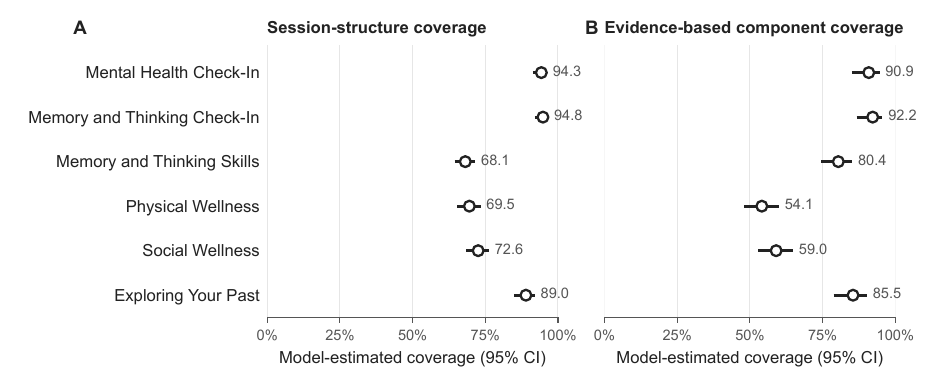}
  \caption{\textbf{Fidelity coverage by session type (449 sessions).} Model-estimated (A)~session-structure coverage and (B)~evidence-based component coverage, with 95\% confidence intervals from binomial mixed models.}
  \Description{Two dot-and-interval panels of coverage percentages for six session types; evidence-based coverage is lowest in Physical and Social Wellness.}
  \label{fig:care_intervention_coverage}
\end{figure}

Among 449 sessions included in the fidelity analysis, both \textit{session-structure coverage} and \textit{evidence-based component coverage} varied substantially by session type (session structure: $\chi^2(5)=258.97$, $p<0.001$; evidence-based components: $\chi^2(5)=158.94$, $p<0.001$; Figure~\ref{fig:care_intervention_coverage}). Estimated session-structure coverage was high for Memory and Thinking Check-In (94.8\%), Mental Health Check-In (94.3\%), and Exploring Your Past (89.0\%), but lower for Memory and Thinking Skills (68.1\%), Physical Wellness (69.5\%), and Social Wellness (72.6\%). Evidence-based component coverage showed a similar pattern, ranging from 90.9\% to 92.2\% in the two check-in sessions and 85.5\% in Exploring Your Past, compared with 80.4\% in Memory and Thinking Skills, 59.0\% in Social Wellness, and 54.1\% in Physical Wellness. Full confidence intervals and pairwise contrasts are reported in Appendix \ref{app:fidelity-estimates}.

These findings indicate that ARC more consistently delivered the intended structure and evidence-based components of the check-in and reminiscence sessions than those of Memory and Thinking Skills, Physical Wellness, and Social Wellness. In particular, Physical and Social Wellness showed substantial omissions of required evidence-based components.

\subsubsection{Off-Protocol Drift Varied Across Session Types}
Exploratory analyses indicated that drift differed by session type, $\chi^2(5)=73.70$, $p<0.001$, with the highest model-estimated rate in Physical Wellness (12.7\%, 95\% CI [7.9, 20.0]) and the lowest in Exploring Your Past (0.4\%, 95\% CI [0.2, 1.0]). Across 449 sessions, 455 of 4,338 assistant turns (10.5\%) were labeled as drift. Functional and boundary deviations were the most common types (Appendix~\ref{app:drift}). Given the lower reliability of drift annotations, these findings remain exploratory (Appendix~\ref{app:llm-validation}).

We also examined when detected drift occurred (Figure~\ref{fig:drift-temporal}). In five of six session types, drift-labeled turns were less frequent early in conversations and peaked later. This pattern was most pronounced in Physical Wellness, increasing from 3\% in the first tenth of conversational progress to 38\% (95\% CI [25, 50]) at 60-70\%. Temporal patterns were descriptive and were not formally tested.

Taken together, these findings show that coherent conversational delivery did not ensure complete delivery of the intended intervention. Coverage analyses identified omissions of required session-structure and evidence-based components, while exploratory drift analyses revealed additional departures from the intended process, particularly in Physical Wellness and later stages of several session types.

\subsubsection{Participant-Specific Contextual Reuse Was Not Significantly Associated with Fidelity}
As a follow-up exploratory analysis, we examined whether stronger participant-specific contextual reuse was associated with the two primary coverage measures. We found no statistically significant association with session-structure coverage (OR per 0.01 increase $=0.979$, 95\% CI [0.952, 1.008], $p=0.158$) or evidence-based intervention component coverage (OR $=0.976$, 95\% CI [0.938, 1.016], $p=0.239$). Results were similar after adjusting for conversation length. The secondary drift analysis also yielded no statistically significant association, but is not interpreted further given its lower annotation agreement (Appendix~\ref{app:personalization-fidelity}).

\subsection{Exploratory RQ3: Context-Dependent Perceptions of Appropriate Data Sharing and Use}
\subsubsection{Comfort Depended on Both the Data Type and Who Would Use or Receive It}
\begin{figure}[tbp]
  \centering
  \includegraphics[width=4.15in]{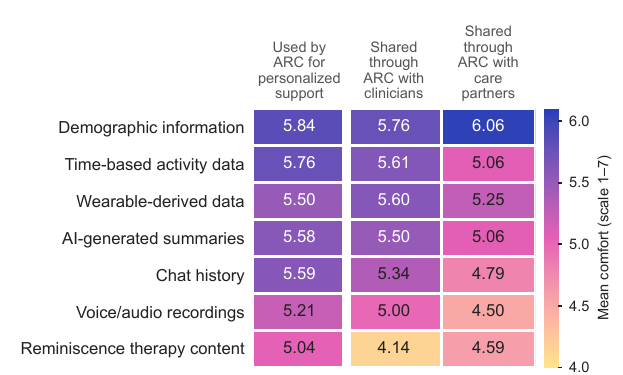}
  \caption{\textbf{Comfort with personal-data use and sharing (exploratory RQ3; $N=80$).} Mean comfort (1--7) by data type and information flow: use by ARC for personalization (Day~3) or sharing through ARC with clinicians or care partners (Day~2).}
  \Description{Heatmap of mean comfort for seven data types under three information flows.}
  \label{fig:rq3_data_sharing_comfort}
\end{figure}

Participants' comfort level in personal data sharing and use varied according to both the recipient and the type of information involved (Figure~\ref{fig:rq3_data_sharing_comfort}). A $3$ (Recipient) $\times$ $7$ (Data Type) repeated-measures ANOVA showed effects of Recipient, $F(2,158)=6.30$, $p=0.002$, $\eta^2_G=0.013$, and Data Type, $F(6,474)=34.17$, $p<0.001$, $\eta^2_G=0.054$, as well as a Recipient $\times$ Data Type interaction, $F(12,948)=8.19$, $p<0.001$, $\eta^2_G=0.014$. Thus, comfort with a given type of information depended in part on who would use or receive it.

Across data types, participants reported the greatest comfort with information used by the AI assistant ($M=5.50$), followed by clinicians ($M=5.28$) and care partners ($M=5.05$). Comfort was higher for AI use than for sharing with care partners ($p=0.010$, Hedges' $g=0.32$) and clinicians ($p=0.045$, $g=0.17$). Across recipients, demographic information was rated most comfortably ($M=5.89$), whereas reminiscence content ($M=4.59$) and voice/audio data ($M=4.90$) were among the least comfortable. Full pairwise comparisons are reported in Appendix \ref{app:rq3}.

\subsubsection{Participants Were More Comfortable with AI Use Than Sharing with Human Care Stakeholders}
In the planned comparison of the two data-use contexts, participants reported greater comfort with information used by ARC for personalized support ($M=5.50$, $SD=1.43$) than with information shared through ARC with clinicians and care partners ($M=5.16$, $SD=1.67$), $t(79)=-3.27$, $p=0.002$, Cohen's $d_z=0.37$.

Together, these exploratory findings show that older adults' comfort with personal-data flows was context-dependent rather than uniform. Comfort varied with both the information involved and who would use or receive it, underscoring the importance of designing data-use and sharing controls around specific information flows within older adults' direct care ecosystems.

\section{Discussion}
\subsection{Scaling Generative AI-Based Psychosocial and Behavioral Health Interventions Requires Evidence of Intervention Delivery Fidelity}
Older adults sustained repeated, multi-turn interactions with ARC and generally reported favorable usability, usefulness, and intentions to use, consistent with prior work showing that voice interfaces can support repeated engagement in health contexts\cite{berube_voice-based_2021, lima_discovering_2023}. Yet these favorable user experiences coexisted with modest participant-specific contextual reuse and omissions of required intervention components, showing why engagement alone provides incomplete evidence for evaluating generative AI-based conversational interventions.

Evaluations of AI-based health interventions should therefore distinguish \textit{interaction experience} (whether people can and want to use a system), \textit{intervention delivery} (whether required processes occur), and \textit{downstream outcomes} (whether those processes produce benefit). Outcome measures alone cannot distinguish an ineffective intervention from one delivered incompletely. This matters when AI is intended to scale existing interventions: access must be accompanied by evidence of faithful delivery. Recent digital mental health research similarly calls for moving beyond feasibility and usability toward evidence about how technologies work, for whom, and under what conditions they can be responsibly integrated into care\cite{birk_digital_2026, torous_evolving_2025}. Evaluating contextual personalization and fidelity of evidence-based intervention delivery can identify delivery limitations and inform interpretation of outcomes in future clinical evaluations.

\subsection{Incomplete Intervention Delivery Can Remain Hidden in Coherent Conversation}
Our coverage analyses identified omissions of required intervention components, even in otherwise coherent conversations. Evaluating generative conversational interventions therefore requires examining both "delivered" content and "prescribed" components that are missing, such as eliciting motivation, developing an action plan, or practicing strategies. This finding underscores the relevance of treatment-fidelity distinctions between prescribed and proscribed therapeutic behaviors\cite{waltz_testing_1993, hadjistavropoulos_development_2018, zainal_capitalizing_2025} for generative AI-based, multi-turn intervention delivery.

An appropriate next response does not necessarily ensure progress through a structured intervention. Recent multi-turn LLM evaluations report difficulties maintaining dialogue structure and following instructions across turns\cite{laban_llms_2025, li_structflowbench_2025}. Such difficulties could contribute to omissions, although our study did not establish their cause. For AI-based health interventions, this suggests that explicitly tracking intervention progress---what has been completed, what remains, and whether the conversation has departed from the intended process---could help maintain delivery fidelity beyond prompt-based instructions alone.

\subsection{The Appropriate Degree of Personalization Depends on the Intervention}
ARC reflected domain-relevant context more strongly than participant-specific context, with participant-specific contextual reuse varying across session types. Participants nevertheless generally experienced ARC as personalized to their health status, preferences, and abilities. These findings suggest that providing context, observing its reuse, and perceiving an interaction as personalized are related but distinct aspects of personalization\cite{kocaballi_personalization_2019}.

Greater participant-specific contextual reuse, however, may not always indicate better personalization. 
Reminiscence may require individual life details, whereas cognitive strategy training relies more on standardized procedures. Under \textit{flexibility within fidelity}\cite{fonagy_fidelity_2019, zilcha-mano_evidence-based_2022}, adaptations should support intervention goals without removing required components. In our exploratory analysis, greater participant-specific contextual reuse was not associated with lower fidelity, although this finding does not rule out a relationship between personalization and fidelity, especially under stronger forms of adaptation. Therefore, putting flexibility within fidelity into practice requires intervention and system designers to specify \textit{which aspects of the intervention can be tailored} to participants’ needs, preferences, and abilities and \textit{which must remain consistent} to preserve fidelity.

\subsection{Personalization Requires Context-Sensitive Data Governance Within Care Relationships}
Comfort varied by data type and recipient, consistent with prior research on older adults' context-dependent sharing preferences\cite{kim_older_2019, crotty_information_2015}. Participants were generally more comfortable with ARC using information for personalization than with sharing it with clinicians or care partners for personalized support, but this does not establish greater AI trust or privacy protection. Contextual Integrity instead situates these judgments within specific information flows, including information type, recipients, and purpose\cite{nissenbaum_contextual_2019, frik_privacy_2019}.

Agents that provide personalized support and communicate with care stakeholders should therefore provide separate, revisable permissions for AI use and sharing with clinicians or care partners. For example, a participant could allow the system to use sleep data for personalized support but exclude it from care-partner summaries. Systems should explain proposed uses, allow changes, and enforce preferences across interactions without requiring users to review every disclosure. Such designs should then be evaluated on both their usability and whether the system respects users’ data-use and sharing permissions\cite{iravantchi_sok_2025}.

\section{Limitations and Future Work}
First, this three-day formative study did not evaluate clinical outcomes or establish readiness for use with older adults with MCI, the population motivating the intervention. The digitally recruited sample had relatively high technology familiarity, prior conversational-agent experience, and relatively preserved everyday cognitive functioning based on ECog-41 scores. These characteristics limit generalizability, but do not establish how interaction experiences or data-sharing expectations would differ in an MCI population. Longer-term evaluation with older adults with MCI should examine comprehension, support needs, and safety alongside usability and delivery fidelity.

Second, contextual reuse captures only one dimension of personalization. Semantic similarity may not capture adaptation in conversational style, pacing, or scaffolding. The short study also cannot establish whether ARC’s prompt-based use of baseline profiles and prior-session summaries supports adaptation as participants’ needs and goals change. Future work should evaluate longitudinal adaptation and complement contextual-reuse measures with human assessments of whether personalization is appropriate and useful.

Third, our personalization-fidelity analysis was observational and participant-specific contextual reuse was modest, limiting causal inference and conclusions about stronger forms of adaptation. Future work should experimentally vary adaptation while holding required intervention components constant to test its effects on fidelity. 

Fourth, our fidelity measures assess whether required components were delivered, but not delivery quality, treatment receipt (participant understanding), or treatment enactment (application outside sessions), as distinguished in established treatment-fidelity frameworks \cite{bellg_enhancing_2004,borrelli_assessment_2011}. Measurement confidence is limited by LLM-assisted annotation and a small human audit of 12 conversations. Expert adjudication supported protocol interpretation but involved the first author, who helped develop the protocols and coding manual. Drift annotations were less reliable than coverage measures and are therefore treated as exploratory. In-context examples, selected as clear illustrations of the coding scheme, were excluded from fidelity analyses. Future work should expand independent validation and assess treatment receipt and enactment, potentially combining self-reports with sensing of everyday behaviors.

Finally, RQ3 measured self-reported comfort with hypothetical information flows, not actual sharing decisions or the reasons behind participants’ judgments. These ratings should not be interpreted as general privacy norms or evidence of greater trust in AI. Qualitative and longitudinal studies should examine how participants explain these expectations and whether they change when information is actually used or shared within care relationships.

\section{Conclusion}
Generative AI offers opportunities to expand access to psychosocial and behavioral interventions, but realizing this potential requires evidence that systems deliver the intended intervention processes. Our formative deployment and evaluation of ARC supports the feasibility of conversational delivery of multidomain lifestyle interventions in this sample of older adults, while revealing delivery gaps that favorable user experiences and perceived personalization alone do not capture. The challenge is therefore to personalize conversational interventions in ways that support intervention goals while respecting participants’ expectations for personal-data use and sharing. Longer-term studies with older adults with MCI are needed to assess whether faithful delivery can be sustained and whether these interventions produce clinical benefit. For ubiquitous computing research, a further opportunity is to connect conversational delivery with everyday enactment, investigating how sensing between sessions could help assess whether participants put intervention strategies into practice.

\begin{acks}
This work was supported by the National Institute on Aging (Grant No. P01 AG073090), the National Science Foundation (Award \#2534209), and Microsoft Agentic AI Research and Innovation (AARI).
Any opinions, findings, conclusions, or recommendations expressed in this material are those of the authors and do not necessarily reflect the views of the funders.
We thank the study participants for their time and contributions, and the reviewers for their constructive feedback. We are grateful to Dimitris Kiosses, Tan Gemicioglu, Maria Teresa Parreira, and Dominik Meier for their feedback on the system prototypes; 
Cristian Danescu-Niculescu-Mizil, Anne Wu, Jerad Moxley,
and Marco Ceruso for their input on the analytical methods;
and Yan Shvartzshnaider and Helen Nissenbaum for their insights into the interpretation of the Privacy Cards findings.
\end{acks}

\bibliographystyle{ACM-Reference-Format}
\bibliography{ref_main, myzotero}

\appendix
\newpage

\renewcommand{\topfraction}{0.9}\renewcommand{\bottomfraction}{0.8}\renewcommand{\textfraction}{0.07}\renewcommand{\floatpagefraction}{0.75}
\setcounter{topnumber}{3}\setcounter{bottomnumber}{2}\setcounter{totalnumber}{4}

\section{Participant Characteristics and Analytic Samples}
\label{app:participant_characteristics}
Table~\ref{tab:table1_demographics} summarizes participants' demographic characteristics and Table~\ref{tab:baseline-health-digital} their baseline health, everyday cognition, and digital proficiency ($N=80$).

\begin{table}[!ht]
\centering
\small
\caption{Participant demographic characteristics ($N=80$). Values are $n$ (\%) unless noted.}
\label{tab:table1_demographics}
\begin{tabular}[t]{@{}lr@{}}
\toprule
Characteristic & Value \\
\midrule
Age, years, $M$ ($SD$) [range] & 70.49 (3.85) [65--80] \\
\addlinespace \textit{Gender} & \\
\quad Female & 54 (67.5\%) \\
\quad Male & 26 (32.5\%) \\
\addlinespace \textit{Race} & \\
\quad White & 70 (87.5\%) \\
\quad Black or African American & 6 (7.5\%) \\
\quad Asian & 2 (2.5\%) \\
\quad White and American Indian/Alaska Native & 1 (1.2\%) \\
\quad White and Black or African American & 1 (1.2\%) \\
\addlinespace \textit{Employment} & \\
\quad Retired & 41 (51.2\%) \\
\quad Working part-time & 20 (25.0\%) \\
\quad Working full-time & 16 (20.0\%) \\
\quad Other & 2 (2.5\%) \\
\quad Unemployed and looking for work & 1 (1.2\%) \\
\bottomrule
\end{tabular}\hspace{2.5em}\begin{tabular}[t]{@{}lr@{}}
\toprule
Characteristic & Value \\
\midrule
\textit{Income} & \\
\quad Less than \$25,000 & 3 (3.8\%) \\
\quad \$25,000--\$49,999 & 22 (27.5\%) \\
\quad \$50,000--\$74,999 & 18 (22.5\%) \\
\quad \$75,000--\$99,999 & 16 (20.0\%) \\
\quad \$100,000--\$149,999 & 14 (17.5\%) \\
\quad \$150,000 or more & 6 (7.5\%) \\
\quad Prefer not to say & 1 (1.2\%) \\
\addlinespace \textit{Education} & \\
\quad Bachelor’s degree & 29 (36.2\%) \\
\quad Graduate or professional degree & 21 (26.2\%) \\
\quad Some college, no degree & 14 (17.5\%) \\
\quad Associate or technical degree & 12 (15.0\%) \\
\quad High school diploma or GED & 3 (3.8\%) \\
\quad Some high school or less & 1 (1.2\%) \\
\addlinespace \textit{Marital Status} & \\
\quad Married & 45 (56.2\%) \\
\quad Divorced/Separated & 16 (20.0\%) \\
\quad Widowed & 13 (16.2\%) \\
\quad Living with a partner & 4 (5.0\%) \\
\quad Never been married & 2 (2.5\%) \\
\bottomrule
\end{tabular}
\par\smallskip\raggedright\footnotesize Race: participants could select more than one category; combinations are listed separately.
\end{table}

\begin{table}[!htbp]
\centering
\small
\caption{Baseline health, everyday cognition, and digital proficiency ($N=80$).}
\label{tab:baseline-health-digital}
\begin{tabular}{@{}lcccc@{}}
\toprule
Measure & Possible range & $M$ ($SD$) & Median & Observed range \\
\midrule
\multicolumn{5}{@{}l}{\textit{Mental health (higher = more symptoms)}} \\
Depressive symptoms (GDS-15) & 0--15 & 2.85 (3.22) & 2 & 0--14 \\
Anxiety symptoms (GAS-10) & 0--30 & 3.62 (3.92) & 2 & 0--18 \\
\quad GDS-15 $\geq$ 5 (screening cutoff), $n$ (\%) & & \multicolumn{3}{l}{18 (22.5\%)} \\
\addlinespace\multicolumn{5}{@{}l}{\textit{Everyday cognition, ECog-41 (mean item rating; higher = more perceived decline)}} \\
Total & 1--4 & 1.35 (0.39) & 1.21 & 1.00--2.56 \\
Memory & 1--4 & 1.54 (0.54) & 1.33 & 1.00--3.25 \\
Language & 1--4 & 1.36 (0.47) & 1.22 & 1.00--3.00 \\
Visuospatial & 1--4 & 1.20 (0.33) & 1.00 & 1.00--2.33 \\
Planning & 1--4 & 1.19 (0.37) & 1.00 & 1.00--2.80 \\
Organization & 1--4 & 1.31 (0.49) & 1.00 & 1.00--3.00 \\
Divided attention & 1--4 & 1.43 (0.59) & 1.25 & 1.00--3.25 \\
\addlinespace\multicolumn{5}{@{}l}{\textit{Mobile device proficiency, MDPQ-16 (subscale mean; higher = more proficient)}} \\
Communication & 1--5 & 4.88 (0.32) & 5.00 & 3.00--5.00 \\
Internet & 1--5 & 4.85 (0.40) & 5.00 & 3.00--5.00 \\
Device operation & 1--5 & 4.76 (0.45) & 5.00 & 3.00--5.00 \\
Entertainment & 1--5 & 4.74 (0.63) & 5.00 & 1.00--5.00 \\
Troubleshooting and software & 1--5 & 4.59 (0.74) & 5.00 & 1.00--5.00 \\
Privacy & 1--5 & 4.58 (0.60) & 5.00 & 2.50--5.00 \\
Calendar & 1--5 & 4.58 (0.96) & 5.00 & 1.00--5.00 \\
Data and file storage & 1--5 & 4.14 (0.93) & 4.00 & 1.00--5.00 \\
\bottomrule
\end{tabular}
\par\smallskip\raggedright\footnotesize \textit{Devices used regularly, $n$ (\%):} Smartphone 79 (98.8\%); Laptop or desktop computer 75 (93.8\%); Tablet 69 (86.2\%); Smart speaker 48 (60.0\%); Smartwatch 43 (53.8\%); Home security system 33 (41.2\%); Smart home controls 29 (36.2\%); Smart ring 1 (1.2\%).\\ \textit{Conversational agents used before, $n$ (\%):} ChatGPT 71 (88.8\%); Siri 53 (66.2\%); Alexa 53 (66.2\%); Retail/shopping chatbot 37 (46.2\%); Google Bard/Gemini 35 (43.8\%); Google Assistant 28 (35.0\%); Banking chatbot 12 (15.0\%). Multiple selections allowed.
\end{table}

\subsection{Analytic Samples}\label{app:samples}
Engagement, perceived acceptability, and contextual-reuse analyses used all 473 retained sessions from the 80 participants who completed all three study days. Fidelity analyses used the subset of these sessions with automated fidelity labels. Of the 473 sessions, 23 had served as in-context examples for the automated annotator and therefore had no automated labels, and one was excluded because of a technical glitch, leaving 449 sessions from 80 participants. The same 449 sessions were used for the session-type fidelity models, the drift analyses, and the personalization--fidelity models.

\section{Representative Voice-Based Conversational Systems}
\label{app:voice-systems}
\begin{figure}[htbp]
    \centering
    \includegraphics[width=0.83\linewidth]{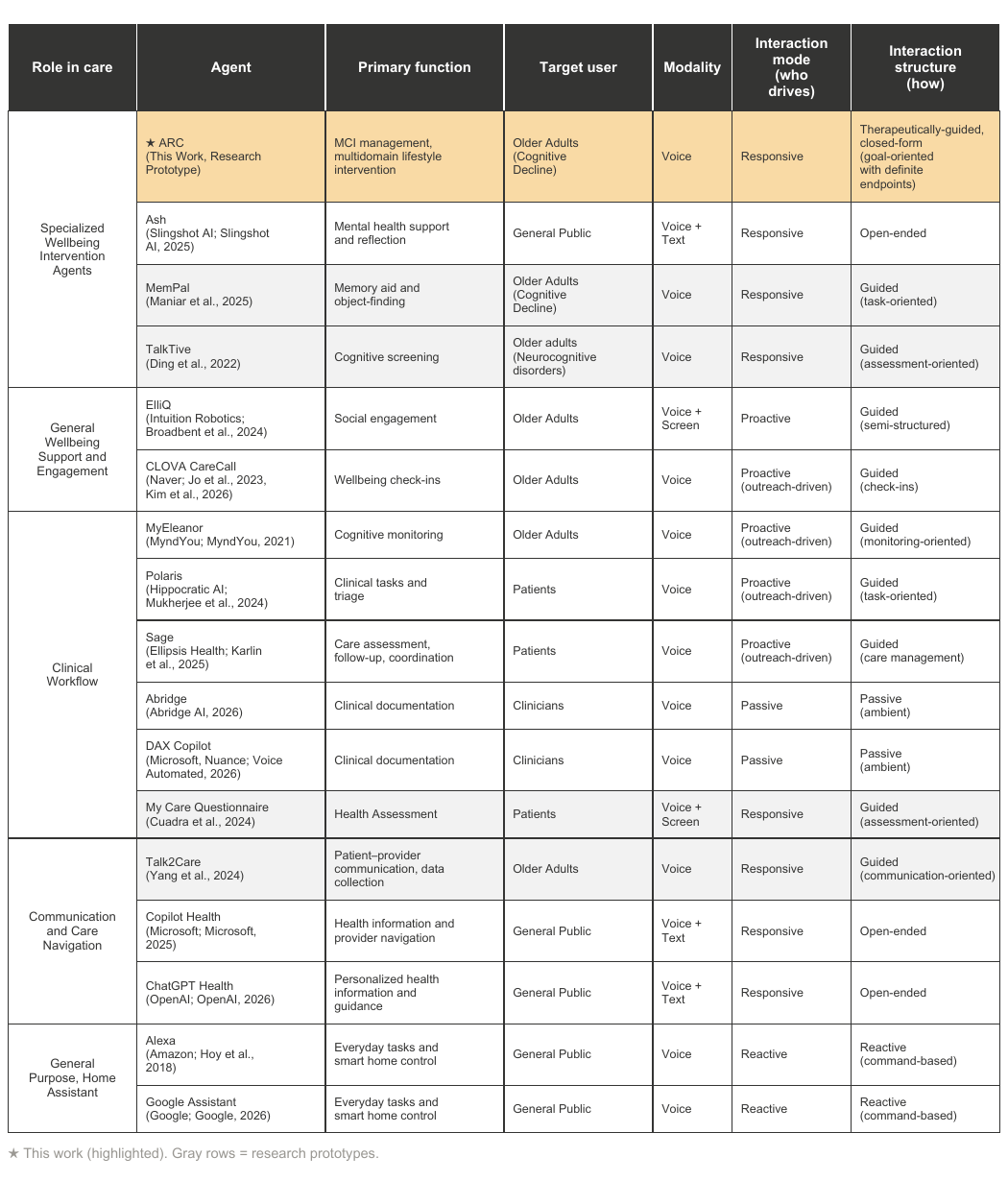}
    \caption{\textbf{Voice-based conversational systems for health and well-being.} Representative systems organized by care role, interaction mode (who leads), and interaction structure. $\bigstar$~marks ARC; gray rows indicate research prototypes.}
  \Description{Table-style figure comparing eighteen voice-based systems; ARC is highlighted at the top.}
  \label{fig:lit}
\end{figure}

Figure~\ref{fig:lit} compares representative voice-based conversational systems by their role in care, interaction mode, and interaction structure.

\section{Evidence-Based Approaches Underlying ARC Sessions}
\label{evidence-based_intervention_principles}
\begin{figure}[htbp]
    \centering
    \includegraphics[width=\linewidth]{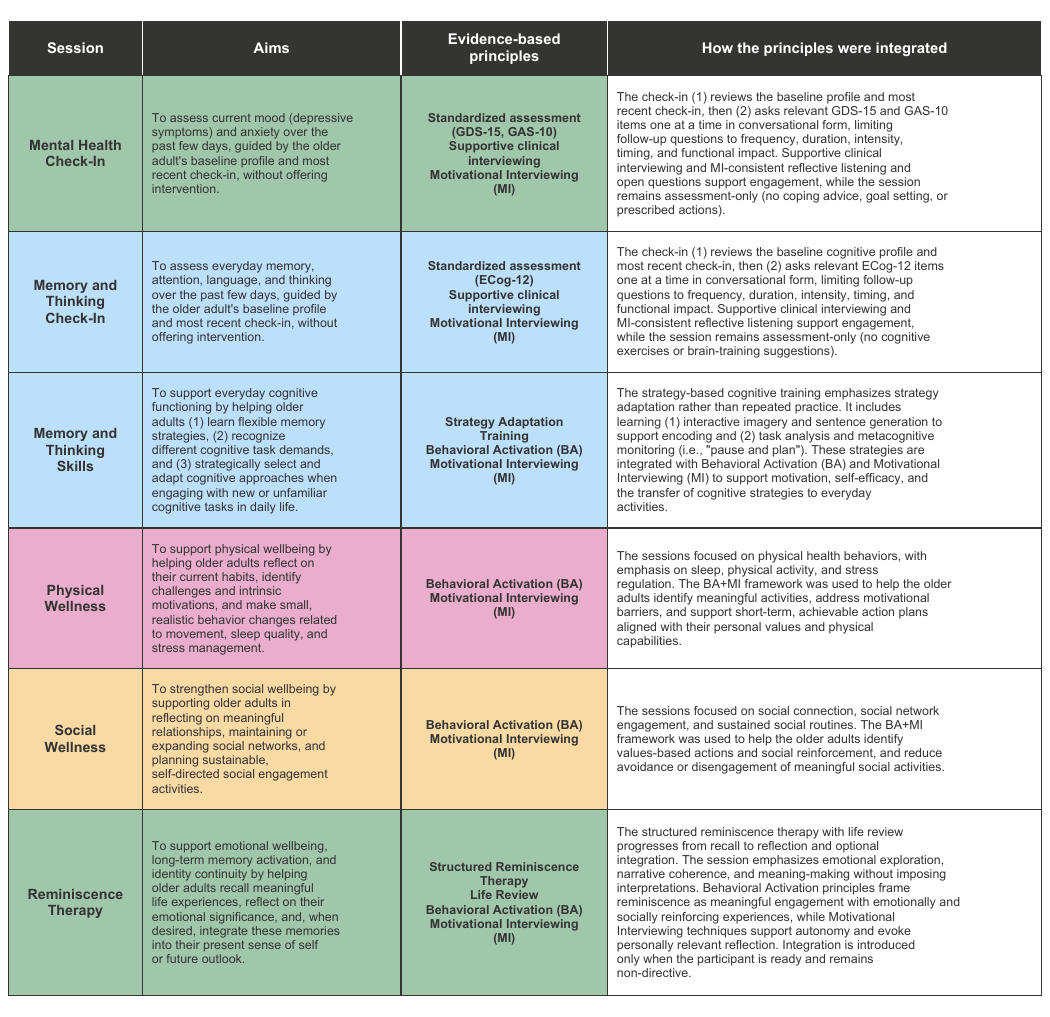}
    \caption{\textbf{Evidence-based principles in ARC's sessions.} Aims, supporting evidence-based principles, and their implementation in conversation for the two health check-ins and the four intervention sessions. In the check-ins, review and structured assessment were scored as evidence-based components.}
  \Description{Table-style figure with one row per session (two check-ins and four intervention sessions): aims, evidence-based principles, and how they were integrated.}
  \label{fig:evidence-based_table}
\end{figure}

Figure~\ref{fig:evidence-based_table} summarizes, for each check-in and intervention session, its aim, the evidence-based principles it draws on, and how they were integrated into the conversation.

\FloatBarrier
\section{RQ1: Engagement and Technology Acceptance}
\label{app:rq1}
Table~\ref{tab:engagement} reports engagement by session type. Table~\ref{tab:tam-days} reports technology acceptance by study day with internal consistency (Cronbach's $\alpha$), and Table~\ref{tab:tam-correlations} reports correlations among the constructs on each day ($N=80$).

\begin{table}[!htbp]
\centering
\small
\caption{Engagement by session type (473 sessions). Duration excludes one Social Wellness session with invalid timestamps ($n=472$).}
\label{tab:engagement}
\resizebox{\linewidth}{!}{%
\begin{tabular}{@{}lccccc@{}}
\toprule
Session type & $n$ & Turns, $M$ ($SD$) & Turns, Mdn [IQR] & Minutes, $M$ ($SD$) & Minutes, Mdn [IQR] \\
\midrule
Mental Health Check-In & 76 & 16.7 (5.5) & 16 [12--20] & 5.3 (2.4) & 4.9 [3.4--6.7] \\
Memory and Thinking Check-In & 78 & 15.3 (4.0) & 14 [12--18] & 5.1 (2.1) & 5.1 [3.3--6.1] \\
Memory and Thinking Skills & 80 & 25.0 (7.5) & 24 [20--28] & 7.2 (2.4) & 7.1 [5.9--8.7] \\
Physical Wellness & 80 & 20.0 (7.3) & 20 [16--23] & 5.5 (2.1) & 5.3 [4.3--6.5] \\
Social Wellness & 79 & 18.0 (5.9) & 16 [14--22] & 5.9 (2.3) & 5.9 [4.8--6.8] \\
Exploring Your Past & 80 & 21.2 (6.9) & 20 [18--24] & 7.1 (2.5) & 6.8 [5.7--8.3] \\
\midrule
All sessions & 473 & 19.4 (7.0) & 18 [14--23] & 6.0 (2.4) & 5.9 [4.3--7.2] \\
\bottomrule
\end{tabular}}
\end{table}

\begin{table}[!htbp]
\centering
\small
\caption{Technology acceptance by day ($N=80$): $M$ ($SD$) on a 1--5 scale and Cronbach's $\alpha$ for Days 1/2/3.}
\label{tab:tam-days}
\begin{tabular}{@{}lcccc@{}}
\toprule
Construct & Day 1 & Day 2 & Day 3 & $\alpha$ (Day 1/2/3) \\
\midrule
Perceived Usefulness & 3.63 (0.72) & 3.93 (0.73) & 3.84 (0.77) & 0.93 / 0.95 / 0.96 \\
Perceived Ease of Use & 3.43 (0.62) & 3.42 (0.63) & 3.19 (0.59) & 0.89 / 0.89 / 0.92 \\
Attitude towards Technology & 3.27 (0.73) & 3.77 (0.84) & 3.81 (0.86) & 0.94 / 0.94 / 0.94 \\
Behavioral Intention & 3.74 (0.82) & 3.77 (1.00) & 3.83 (0.91) & 0.96 / 0.98 / 0.98 \\
\bottomrule
\end{tabular}
\end{table}

\begin{table}[!htbp]
\centering
\small
\caption{Pearson correlations among technology acceptance constructs by day ($N=80$). PU = perceived usefulness; PEOU = perceived ease of use; ATT = attitude; BI = behavioral intention.}
\label{tab:tam-correlations}
\begin{tabular}{@{}lccc@{}}
\toprule
Construct pair & Day 1 & Day 2 & Day 3 \\
\midrule
PU -- PEOU & 0.48 & 0.56 & 0.48 \\
PU -- ATT & 0.70 & 0.75 & 0.76 \\
PU -- BI & 0.82 & 0.79 & 0.76 \\
PEOU -- ATT & 0.70 & 0.79 & 0.72 \\
PEOU -- BI & 0.50 & 0.64 & 0.46 \\
ATT -- BI & 0.69 & 0.82 & 0.76 \\
\bottomrule
\end{tabular}
\end{table}

\section{RQ2a: Contextual Personalization}
\label{personalization_contexts_appendix}
\begin{figure}[htbp]
    \centering
    \includegraphics[width=\linewidth]{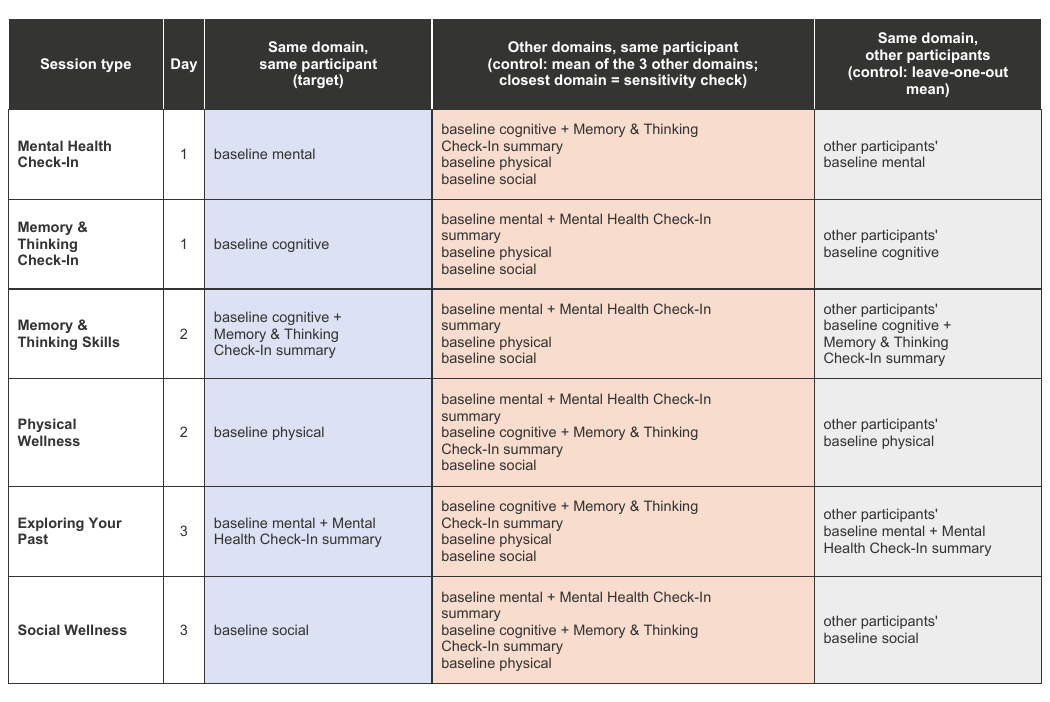}
    \caption{\textbf{Context conditions for contextual-reuse analysis.} Each session type is compared against its target context (same participant, same domain) and two controls: the participant's other-domain contexts (mean across three; closest domain in a sensitivity check) and other participants' same-domain contexts (leave-one-out mean).}
  \Description{Table-style figure listing, for each session type, the contexts used in the target condition and the two control conditions.}
  \label{fig:personalization_contexts}
\end{figure}

Table~\ref{tab:rq2a-session-types} reports contextual reuse for each session type. Responses were more similar to the participant's own domain-matched context than to the mean of the three non-target domains in every session type. Against the single closest non-target domain, the difference remained in five of six session types; in Physical Wellness, responses were about as similar to the closest non-target context as to the target context. Participant-specific effects were small and survived FDR correction only for Memory and Thinking Check-In, Physical Wellness, and Exploring Your Past.

\begin{table}[!htbp]
\centering
\small
\caption{Contextual reuse by session type. $\Delta$ = paired mean difference in session-level cosine similarity (target $-$ control; $t$-based 95\% CI); $d$ = mean difference divided by the mean condition SD. Tests are two-tailed Wilcoxon signed-rank; $q$ = BH-FDR-adjusted $p$ within each six-test comparison family. The closest-domain sensitivity check reports raw $p$; the primary pooled test is not FDR-adjusted. Each participant contributes at most one session per type.}
\label{tab:rq2a-session-types}
\resizebox{\linewidth}{!}{%
\begin{tabular}{@{}lcccccccccc@{}}
\toprule
 & Target & \multicolumn{3}{c}{vs.\ other domains (mean of 3)} & \multicolumn{3}{c}{vs.\ closest other domain} & \multicolumn{3}{c}{vs.\ other participants} \\
\cmidrule(lr){3-5}\cmidrule(lr){6-8}\cmidrule(l){9-11}
Session type ($n$) & $M$ & $\Delta$ [95\% CI] & $d$ & $q$ & $\Delta$ [95\% CI] & $d$ & $p$ & $\Delta$ [95\% CI] & $d$ & $q$ \\
\midrule
Mental Health Check-In (76) & 0.280 & 0.093 [0.084, 0.102] & 2.15 & $<$0.001 & 0.041 [0.030, 0.051] & 0.88 & $<$0.001 & 0.007 [0.002, 0.013] & 0.15 & 0.053 \\
Memory and Thinking Check-In (78) & 0.385 & 0.168 [0.155, 0.182] & 3.26 & $<$0.001 & 0.105 [0.091, 0.119] & 1.91 & $<$0.001 & 0.007 [0.002, 0.013] & 0.11 & 0.048 \\
Memory and Thinking Skills (80) & 0.348 & 0.202 [0.191, 0.213] & 4.78 & $<$0.001 & 0.154 [0.142, 0.166] & 3.37 & $<$0.001 & 0.003 [-0.003, 0.009] & 0.06 & 0.536 \\
Physical Wellness (80) & 0.289 & 0.079 [0.065, 0.093] & 1.32 & $<$0.001 & 0.011 [-0.006, 0.028] & 0.16 & 0.126 & 0.008 [-0.000, 0.017] & 0.13 & 0.048 \\
Social Wellness (79) & 0.277 & 0.127 [0.114, 0.140] & 2.10 & $<$0.001 & 0.060 [0.046, 0.074] & 0.89 & $<$0.001 & 0.004 [-0.004, 0.011] & 0.05 & 0.518 \\
Exploring Your Past (80) & 0.203 & 0.079 [0.070, 0.088] & 1.58 & $<$0.001 & 0.019 [0.009, 0.029] & 0.35 & $<$0.001 & 0.011 [0.002, 0.019] & 0.20 & 0.048 \\
\midrule
All sessions (473) & 0.297 & 0.125 [0.119, 0.131] & 1.81 & $<$0.001 & 0.065 [0.058, 0.072] & 0.88 & $<$0.001 & 0.007 [0.004, 0.010] & 0.08 & $<$0.001 \\
\bottomrule
\end{tabular}}
\end{table}

Table~\ref{tab:personalization-days} reports perceived personalization by study day.

\begin{table}[!htbp]
\centering
\small
\caption{Perceived personalization by dimension and day ($N=80$): $M$ ($SD$) on a 1--5 scale.}
\label{tab:personalization-days}
\begin{tabular}{@{}lccc@{}}
\toprule
Dimension & Day 1 & Day 2 & Day 3 \\
\midrule
Tailored to health status & 4.17 (0.85) & 3.86 (1.08) & 3.86 (0.96) \\
Tailored to abilities & 3.95 (1.08) & 3.85 (1.03) & 3.94 (1.06) \\
Tailored to preferences & 3.86 (0.87) & 3.71 (1.05) & 3.79 (1.05) \\
\bottomrule
\end{tabular}
\par\smallskip\raggedright\footnotesize 3 (dimension) $\times$ 3 (day) repeated-measures ANOVA: dimension $F(2,158)=4.56$, $p=0.012$, $\eta^2_G=0.006$; day $F(2,158)=2.13$, $p=0.122$, $\eta^2_G=0.006$; interaction $F(4,316)=1.93$, $p=0.135$, $\eta^2_G=0.003$. Bonferroni-corrected comparisons: ability vs.\ health status, $p=1.000$, $g=-0.07$; ability vs.\ preferences, $p=0.184$, $g=0.15$; health status vs.\ preferences, $p=0.002$, $g=0.23$.
\end{table}

\FloatBarrier
\section{RQ2b: Fidelity of Intervention Delivery}
\subsection{Coverage by Session Type}
\label{app:fidelity-estimates}
\begin{figure*}[htbp]
    \centering
    \includegraphics[width=\linewidth]{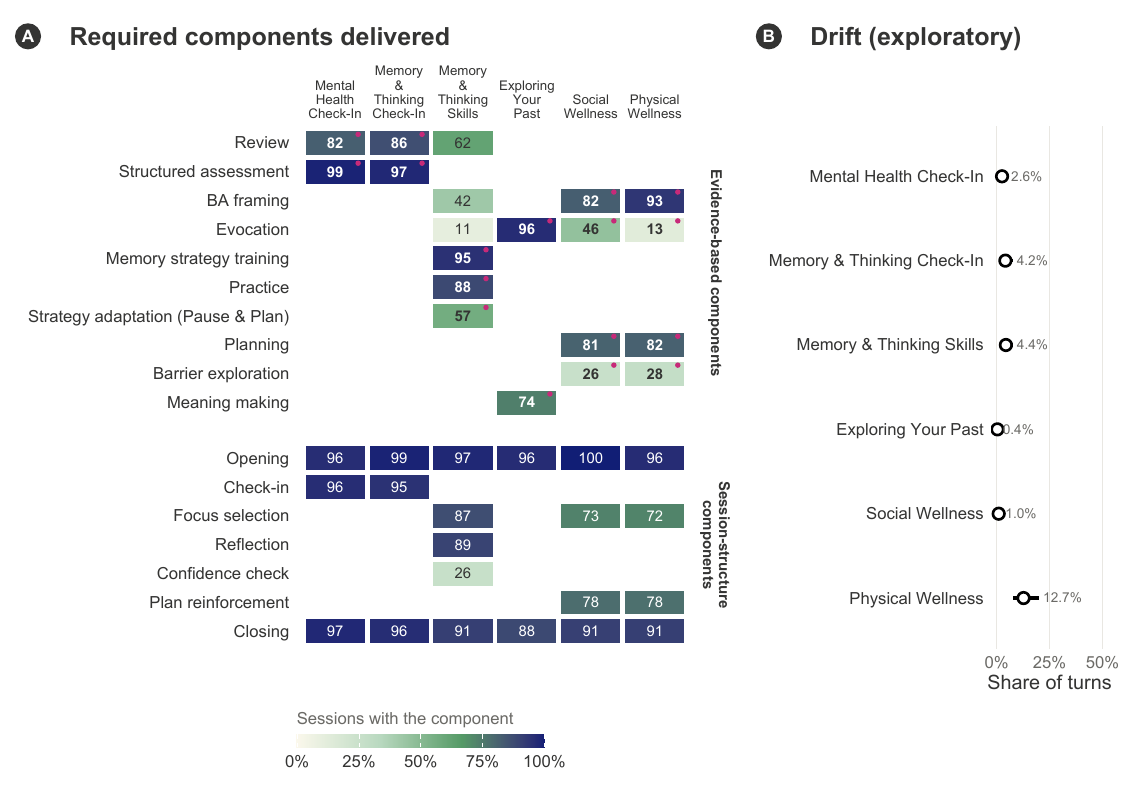}
    \caption{\textbf{Component coverage and exploratory drift (449 sessions).} (A)~Proportion of sessions containing each required component, grouped by component type; magenta dots identify components counted toward evidence-based coverage. (B)~Model-estimated drift rate for a typical session, with 95\% CI.}
  \Description{Heatmap of component delivery by session type next to a dot plot of drift rates.}
  \label{fig:fidelity-diagnostics}
\end{figure*}

Table~\ref{tab:fidelity-coverage} reports model-estimated and observed coverage by session type, and Figure~\ref{fig:fidelity-diagnostics} shows which required components were delivered. Pairwise comparisons used Tukey-adjusted contrasts (Table~\ref{tab:fidelity-pairwise}); odds ratios (OR) are reported for significant contrasts.

\textit{Session-structure coverage.} Both check-ins had higher coverage than Memory and Thinking Skills, Physical Wellness, and Social Wellness (OR = 6.2--8.6, all $p<0.001$). Exploring Your Past also exceeded these three sessions (OR = 3.1--3.8, all $p<0.001$). The check-ins did not differ from each other or from Exploring Your Past, and Memory and Thinking Skills, Physical Wellness, and Social Wellness did not differ from one another.

\textit{Evidence-based component coverage.} Both check-ins had higher coverage than Physical and Social Wellness (OR = 6.9--10.1, all $p<0.001$), and the Memory and Thinking Check-In also exceeded Memory and Thinking Skills (OR = 2.9, $p=0.026$). Memory and Thinking Skills and Exploring Your Past each exceeded Physical Wellness (OR = 3.5 and 5.0) and Social Wellness (OR = 2.8 and 4.1; all $p<0.001$). Physical and Social Wellness did not differ.

\begin{table}[!htbp]
\centering
\small
\caption{Session-structure and evidence-based component coverage by session type (449 sessions; 80 participants). Model-estimated percentages (95\% CI) from binomial-logit mixed models with session type fixed and participant random intercepts; observed percentages are descriptive session-level means.}
\label{tab:fidelity-coverage}
\resizebox{\linewidth}{!}{%
\begin{tabular}{@{}lccccc@{}}
\toprule
 & & \multicolumn{2}{c}{Session-structure coverage} & \multicolumn{2}{c}{Evidence-based component coverage} \\
\cmidrule(lr){3-4}\cmidrule(l){5-6}
Session type & Sessions & Model [95\% CI] & Observed & Model [95\% CI] & Observed \\
\midrule
Mental Health Check-In & 73 & 94.3 [91.4, 96.2] & 94.0 & 90.9 [85.0, 94.6] & 90.4 \\
Memory and Thinking Check-In & 74 & 94.8 [92.1, 96.7] & 94.6 & 92.2 [86.7, 95.6] & 91.9 \\
Memory and Thinking Skills & 76 & 68.1 [64.5, 71.6] & 67.7 & 80.4 [74.5, 85.2] & 79.8 \\
Physical Wellness & 76 & 69.5 [65.4, 73.4] & 69.1 & 54.1 [48.1, 60.0] & 53.9 \\
Social Wellness & 74 & 72.6 [68.5, 76.3] & 72.1 & 59.0 [52.9, 64.8] & 58.8 \\
Exploring Your Past & 76 & 89.0 [84.9, 92.1] & 88.5 & 85.5 [78.8, 90.3] & 84.9 \\
\bottomrule
\end{tabular}}
\par\smallskip\raggedright\footnotesize Omnibus likelihood-ratio tests of session type: session-structure coverage $\chi^2(5)=258.97$, $p<0.001$; evidence-based component coverage $\chi^2(5)=158.94$, $p<0.001$.
\end{table}

\begin{table}[!htbp]
\centering
\small
\caption{Pairwise session-type comparisons of coverage and exploratory drift: odds ratios and Tukey-adjusted $p$-values from models in Tables~\ref{tab:fidelity-coverage} and~\ref{tab:drift}. Odds ratios above 1 indicate higher coverage or drift for the first session type.}
\label{tab:fidelity-pairwise}
\resizebox{\linewidth}{!}{%
\begin{tabular}{@{}lccc@{}}
\toprule
Comparison & Session-structure coverage & Evidence-based component coverage & Drift (exploratory) \\
\midrule
MTC vs.\ MH & 1.11 (0.999) & 1.19 (0.998) & 1.64 (0.852) \\
MTC vs.\ MTS & 8.58 ($<$0.001) & 2.89 (0.026) & 0.94 (1.000) \\
MTC vs.\ PW & 8.04 ($<$0.001) & 10.06 ($<$0.001) & 0.30 (0.022) \\
MTC vs.\ EYP & 2.28 (0.056) & 2.02 (0.431) & 10.66 ($<$0.001) \\
MTC vs.\ SW & 6.94 ($<$0.001) & 8.23 ($<$0.001) & 4.24 (0.020) \\
MH vs.\ MTS & 7.70 ($<$0.001) & 2.43 (0.074) & 0.58 (0.757) \\
MH vs.\ PW & 7.21 ($<$0.001) & 8.45 ($<$0.001) & 0.18 ($<$0.001) \\
MH vs.\ EYP & 2.04 (0.122) & 1.70 (0.693) & 6.51 (0.003) \\
MH vs.\ SW & 6.23 ($<$0.001) & 6.91 ($<$0.001) & 2.59 (0.319) \\
MTS vs.\ PW & 0.94 (0.993) & 3.48 ($<$0.001) & 0.32 (0.023) \\
MTS vs.\ EYP & 0.27 ($<$0.001) & 0.70 (0.800) & 11.29 ($<$0.001) \\
MTS vs.\ SW & 0.81 (0.479) & 2.85 ($<$0.001) & 4.49 (0.010) \\
PW vs.\ EYP & 0.28 ($<$0.001) & 0.20 ($<$0.001) & 35.76 ($<$0.001) \\
PW vs.\ SW & 0.86 (0.864) & 0.82 (0.839) & 14.22 ($<$0.001) \\
EYP vs.\ SW & 3.05 ($<$0.001) & 4.08 ($<$0.001) & 0.40 (0.509) \\
\bottomrule
\end{tabular}}
\par\smallskip\raggedright\footnotesize MH = Mental Health Check-In; MTC = Memory and Thinking Check-In; MTS = Memory and Thinking Skills; PW = Physical Wellness; SW = Social Wellness; EYP = Exploring Your Past.
\end{table}

\subsection{Exploratory Drift}
\label{app:drift}
Drift episodes were defined as runs of consecutive assistant turns labeled with any drift subtype (214 episodes in 160 of the 449 sessions; 455 drift turns among 4,338 assistant turns). Table~\ref{tab:drift} reports observed and model-estimated drift by session type. Model estimates are conditional on the random effects being zero, so they describe a typical session and are lower than the pooled observed shares, which are dominated by a minority of sessions with frequent drift. Physical Wellness had a higher drift rate than every other session type (OR = 3.2--35.8, all $p\le 0.023$). Exploring Your Past had a lower drift rate than both check-ins, Memory and Thinking Skills, and Physical Wellness (OR = 6.5--35.8, all $p\le 0.003$) but did not differ from Social Wellness (OR = 2.5, $p=0.509$). Social Wellness had a lower drift rate than the Memory and Thinking Check-In (OR = 4.2, $p=0.020$) and Memory and Thinking Skills (OR = 4.5, $p=0.010$).

\begin{figure*}[htbp]
    \centering
    \includegraphics[width=0.78\linewidth]{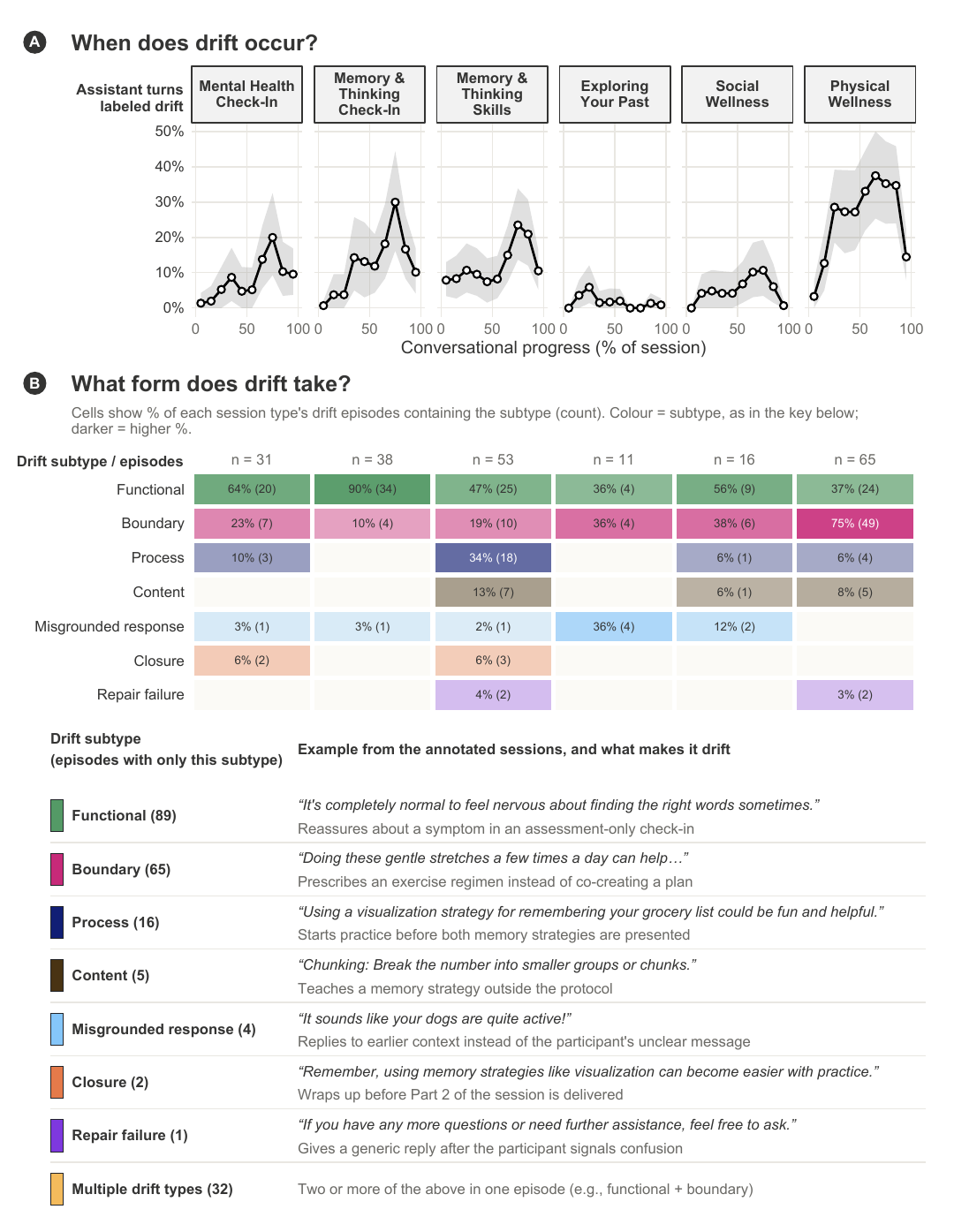}
    \caption{\textbf{Timing and subtypes of detected drift (exploratory).} (A)~Proportion of assistant turns labeled drift across ten bins of conversational progress (95\% participant-bootstrap confidence intervals). (B)~Proportion of drift episodes containing each subtype, by session type (214 episodes total); episodes may contain multiple subtypes. The key gives an example of each subtype.}
  \Description{Line charts of drift share across conversational progress for six session types, and a tile chart of drift subtypes with example quotes.}
  \label{fig:drift-temporal}
\end{figure*}

\textit{When drift occurred.} In five of six session types, drift was rare at the start of the session and most frequent in the later middle of the conversation (Figure~\ref{fig:drift-temporal}A). Drift was most pronounced in Physical Wellness, rising from 3\% of assistant turns in the first decile of the session to a peak of 38\% (95\% CI [25\%, 50\%]) at 60--70\% of conversational progress. Drift peaked at 70--80\% of progress in the Memory and Thinking Check-In (30\% [16\%, 44\%]), Memory and Thinking Skills (24\% [14\%, 34\%]), and the Mental Health Check-In (20\% [9\%, 33\%]). Drift remained at or below 11\% in Social Wellness and 6\% in Exploring Your Past in every decile.

\textit{What form drift took.} Across episodes, the most common subtypes were functional drift (116 of 214 episodes, 54\%) and boundary drift (80, 37\%), followed by process drift (26, 12\%); episodes could contain more than one subtype. The dominant subtype differed by session type (Table~\ref{tab:drift-subtypes}): functional drift predominated in the check-ins, boundary drift in Physical Wellness, and process drift was concentrated in Memory and Thinking Skills.

\begin{table}[!htbp]
\centering
\small
\caption{Drift by session type (exploratory; 449 sessions, 80 participants). Observed = drift-labeled turns / assistant turns. Model = estimated drift share for a typical session (binomial-logit mixed model; session type fixed; participant and observation-level random intercepts set to zero; 95\% CI). These conditional estimates are lower than pooled observed shares because observation-level variance is large. Episodes are runs of consecutive drift-labeled assistant turns.}
\label{tab:drift}
\resizebox{\linewidth}{!}{%
\begin{tabular}{@{}lccccccc@{}}
\toprule
 & & Assistant & Drift & Observed & Model & Sessions & Drift \\
Session type & Sessions & turns & turns & \% & \% [95\% CI] & with drift & episodes \\
\midrule
Mental Health Check-In & 73 & 608 & 48 & 7.9 & 2.6 [1.4, 4.9] & 28 & 31 \\
Memory and Thinking Check-In & 74 & 564 & 65 & 11.5 & 4.2 [2.3, 7.6] & 31 & 38 \\
Memory and Thinking Skills & 76 & 939 & 107 & 11.4 & 4.4 [2.5, 7.6] & 36 & 53 \\
Physical Wellness & 76 & 754 & 184 & 24.4 & 12.7 [7.9, 20.0] & 43 & 65 \\
Social Wellness & 74 & 672 & 35 & 5.2 & 1.0 [0.5, 2.2] & 12 & 16 \\
Exploring Your Past & 76 & 801 & 16 & 2.0 & 0.4 [0.2, 1.0] & 10 & 11 \\
\midrule
All & 449 & 4338 & 455 & 10.5 & & 160 & 214 \\
\bottomrule
\end{tabular}}
\par\smallskip\raggedright\footnotesize Omnibus likelihood-ratio test of session type (full and reduced models with the same random effects): $\chi^2(5)=73.70$, $p<0.001$. The observation-level random effect was added because drift counts were overdispersed under a participant-only model (DHARMa dispersion test $p=0.006$; with the observation-level effect, $p=0.73$).
\end{table}

\begin{table}[!htbp]
\centering
\small
\caption{Drift subtypes by session type (exploratory). Values are the percentage (count) of drift episodes containing each subtype. Episodes may contain multiple subtypes, so percentages across columns may exceed 100\%.}
\label{tab:drift-subtypes}
\begin{tabular}{@{}lcccccc@{}}
\toprule
Subtype & MH & MTC & MTS & PW & SW & EYP \\
\midrule
Episodes & 31 & 38 & 53 & 65 & 16 & 11 \\
\midrule
Functional & 64 (20) & 90 (34) & 47 (25) & 37 (24) & 56 (9) & 36 (4) \\
Boundary & 23 (7) & 10 (4) & 19 (10) & 75 (49) & 38 (6) & 36 (4) \\
Process & 10 (3) & 0 & 34 (18) & 6 (4) & 6 (1) & 0 \\
Content & 0 & 0 & 13 (7) & 8 (5) & 6 (1) & 0 \\
Misgrounded response & 3 (1) & 3 (1) & 2 (1) & 0 & 12 (2) & 36 (4) \\
Closure & 6 (2) & 0 & 6 (3) & 0 & 0 & 0 \\
Repair failure & 0 & 0 & 4 (2) & 3 (2) & 0 & 0 \\
\bottomrule
\end{tabular}
\par\smallskip\raggedright\footnotesize Abbreviations as in Table~\ref{tab:fidelity-pairwise}.
\end{table}

\subsection{Participant-Specific Contextual Reuse and Fidelity}
\label{app:personalization-fidelity}
Table~\ref{tab:personalization-fidelity} reports the full personalization--fidelity models (449 sessions, 80 participants). Each model was a binomial-logit mixed model with participant-specific contextual reuse (centered within session type) and session type as fixed effects and a participant random intercept; sensitivity models additionally adjusted for conversation length. Coverage models showed no overdispersion. Drift counts were overdispersed relative to the binomial distribution (DHARMa dispersion test $p=0.012$), so the drift model additionally included an observation-level random effect; without it, the drift association appeared significant (OR = 1.055, $p=0.006$), an artifact of the underestimated standard errors. No outcome was associated with participant-specific contextual reuse.

\begin{table}[!htbp]
\centering
\small
\caption{Associations between participant-specific contextual reuse and fidelity (449 sessions; 80 participants). Odds ratios per 0.01 increase in reuse (approximately one-third SD) from binomial-logit mixed models, with reuse centered within session type, session type fixed, and participant random intercepts. Wald 95\% CIs; exploratory, unadjusted $p$-values.}
\label{tab:personalization-fidelity}
\begin{tabular}{@{}lcccc@{}}
\toprule
 & \multicolumn{2}{c}{Primary model} & \multicolumn{2}{c}{+ Conversation length} \\
\cmidrule(lr){2-3}\cmidrule(l){4-5}
Outcome & OR [95\% CI] & $p$ & OR [95\% CI] & $p$ \\
\midrule
Session-structure coverage & 0.98 [0.95, 1.01] & 0.158 & 0.98 [0.96, 1.01] & 0.278 \\
Evidence-based component coverage & 0.98 [0.94, 1.02] & 0.239 & 0.98 [0.94, 1.02] & 0.377 \\
Drift rate (exploratory)$^{a}$ & 1.06 [0.98, 1.15] & 0.123 & 1.06 [0.98, 1.15] & 0.126 \\
\bottomrule
\end{tabular}
\par\smallskip\raggedright\footnotesize $^{a}$ Includes an observation-level random effect because drift counts were overdispersed (DHARMa dispersion test $p=0.012$).
\end{table}

\subsection{Validation of LLM-Derived Fidelity Measures}
\label{app:llm-validation}
We assessed whether the session-level fidelity measures computed from LLM labels were reproduced by human annotation of the same sessions, using two comparison sets (Table~\ref{tab:llm-validation}): an audit of 12 previously unseen sessions (two per session type), annotated by a co-author blinded to the LLM labels and adjudicated by the first author without access to the LLM output, and the first author's reference labels for the 36 held-out validation sessions. In the audit, agreement was ICC~$=0.82$ [0.49, 0.95] for session-structure coverage and 0.74 [0.34, 0.91] for evidence-based component coverage, with mean LLM--human differences whose confidence intervals included zero. For drift, the LLM labeled 15 of the 30 turns that the human annotators labeled as drift, and all 15 of its own drift labels fell on turns that the human annotators also labeled as drift. Given the small audit sample (12 sessions), these results are descriptive. Drift agreement on the reference set was weak (ICC~$=0.22$).

\begin{table}[!htbp]
\centering
\small
\caption{Agreement between LLM- and human-derived fidelity measures. The primary audit used 12 sessions unseen during codebook development or in-context prompting, independently annotated by a co-author blinded to LLM labels and adjudicated by the first author without access to LLM output. The 36-session held-out validation set used first-author reference labels. ICC(A,1) = two-way random-effects, absolute-agreement, single-rater coefficient.}
\label{tab:llm-validation}
\resizebox{\linewidth}{!}{%
\begin{tabular}{@{}lcc@{}}
\toprule
Measure & Adjudicated audit (12 sessions) & Reference labels (36 sessions) \\
\midrule
Session-structure coverage: ICC(A,1) [95\% CI] & 0.82 [0.49, 0.95] & 0.67 [0.44, 0.82] \\
\quad Bias, LLM $-$ human [95\% CI] & $-$0.04 [$-$0.10, 0.01] & +0.00 [$-$0.05, 0.05] \\
Evidence-based component coverage: ICC(A,1) & 0.74 [0.34, 0.91] & 0.59 [0.33, 0.77] \\
\quad Bias & $-$0.06 [$-$0.16, 0.04] & +0.01 [$-$0.08, 0.09] \\
Session-structure component present vs omitted: $\kappa$; \% agreement & 0.72 [0.53, 0.89]; 90\% & 0.64 [0.50, 0.75]; 88\% \\
Evidence-based component present vs omitted: $\kappa$; \% agreement & 0.72 [0.26, 0.94]; 88\% & 0.50 [0.26, 0.70]; 80\% \\
Drift rate (exploratory): ICC(A,1) & 0.67 [0.22, 0.89] & 0.22 [$-$0.11, 0.51] \\
\quad Bias & $-$0.09 [$-$0.22, 0.04] & +0.02 [$-$0.06, 0.10] \\
\bottomrule
\end{tabular}}
\end{table}

\FloatBarrier
\section{Exploratory RQ3: Context-Dependent Comfort with Data Use and Sharing}
\label{app:rq3}
Table~\ref{tab:rq3-cells} reports mean comfort for each data type and information flow. Bonferroni-corrected comparisons of the three flows (averaged across data types) showed higher comfort for use by ARC than for sharing through ARC with care partners ($p=0.010$, $g=0.32$) or clinicians ($p=0.045$, $g=0.17$); sharing with care partners and with clinicians did not differ ($p=0.28$). Because use by ARC and sharing through ARC were rated in different Privacy Card scenarios on different days, these differences describe the information flows as wholes rather than an effect of recipient identity. Across recipients, demographic information was rated more comfortable than every other data type (all $p<0.001$, $g=0.33$--$0.98$), and reminiscence content was rated less comfortable than every other type (all $p<0.001$, $g=0.46$--$0.98$) except voice/audio recordings ($p=0.063$). Voice/audio recordings were rated less comfortable than AI-generated summaries, chat history, time-based activity data, and wearable data (all $p\le 0.005$, $g=0.23$--$0.40$). Time-based activity data, wearable data, AI-generated summaries, and chat history did not differ from one another after correction.

\begin{table}[!htbp]
\centering
\small
\caption{Comfort with personal-data use and sharing (exploratory RQ3; $N=80$): $M$ ($SD$) on a 1--7 scale (1 = very uncomfortable; 7 = very comfortable). Ratings reflect hypothetical use or sharing, not actual decisions. ARC use and ARC-mediated sharing were presented in different Privacy Card scenarios on different days.}
\label{tab:rq3-cells}
\resizebox{\linewidth}{!}{%
\begin{tabular}{@{}lcccc@{}}
\toprule
Data type & Used by ARC & Shared with clinicians & Shared with care partners & Mean \\
\midrule
Demographic information & 5.84 (1.19) & 5.76 (1.36) & 6.06 (1.40) & 5.89 \\
Time-based activity data & 5.76 (1.29) & 5.61 (1.50) & 5.06 (1.92) & 5.48 \\
Wearable-derived data & 5.50 (1.57) & 5.60 (1.44) & 5.25 (1.84) & 5.45 \\
AI-generated summaries & 5.58 (1.27) & 5.50 (1.48) & 5.06 (1.87) & 5.38 \\
Chat history & 5.59 (1.28) & 5.34 (1.57) & 4.79 (1.90) & 5.24 \\
Voice/audio recordings & 5.21 (1.60) & 5.00 (1.81) & 4.50 (2.11) & 4.90 \\
Reminiscence therapy content & 5.04 (1.60) & 4.14 (1.95) & 4.59 (2.07) & 4.59 \\
\midrule
All data types & 5.50 & 5.28 & 5.04 & \\
\bottomrule
\end{tabular}}
\par\smallskip\raggedright\footnotesize 3 (information flow) $\times$ 7 (data type) repeated-measures ANOVA: information flow $F(2,158)=6.30$, GG-corrected $p=0.005$, $\eta^2_G=0.013$; data type $F(6,474)=34.17$, GG-corrected $p<0.001$, $\eta^2_G=0.054$; interaction $F(12,948)=8.19$, GG-corrected $p<0.001$, $\eta^2_G=0.014$. Bonferroni-corrected comparisons of information flows: use by ARC vs.\ sharing with care partners: $p=0.010$, $g=0.32$; use by ARC vs.\ sharing with clinicians: $p=0.045$, $g=0.17$; sharing with care partners vs.\ sharing with clinicians: $p=0.275$, $g=0.16$. Planned contrast, use by ARC vs.\ sharing through ARC (averaged over recipients and data types): $M=5.50$ vs.\ $5.16$, $t(79)=3.27$, $p=0.002$, $d_z=0.37$. Pairwise data-type comparisons are reported in the text.
\end{table}

\FloatBarrier
\section{Safety Protocols and Response Specifications}
\label{appendix:safety_protocols}
\subsection{Safety Architecture Overview}
We configured a two-layer safety architecture combining platform-level content filtering with application-level response protocols. Platform-level filtering is designed to detect and block high-risk content before it reaches the application, and application-level guardrails specify how ARC should respond to harmful, sensitive, or out-of-scope inputs. The sections below describe these configurations and instructions; we did not formally evaluate their performance in this study.

\subsection{Platform-Level Content Filtering}
The system relied on Azure OpenAI’s built-in content filtering and abuse monitoring mechanisms. These apply neural classifiers to both user inputs and model outputs to detect harmful content across four primary categories: hate and harassment, sexual content, violence, and self-harm \cite{patrickfarley_content_2026}. Content is assessed across multiple severity levels, and inputs rated high-risk are blocked before reaching the application layer. This filtering operates independently of application logic.

\subsection{Application-Level Response Principles}
For inputs not addressed by automated filtering, the system prompts specified response protocols guided by the following principles:
\begin{itemize}
    \item \textbf{Non-amplification:} Avoid affirming, repeating, or escalating harmful language, intent, or misinformation.
    \item \textbf{Role clarity:} Explicitly state the system’s role as an AI Health Assistant Tool, not a human clinician, caregiver, or advisor.
    \item \textbf{Emotional acknowledgment:} Recognize user emotions or distress in a non-judgmental manner without endorsing harmful beliefs or actions.
    \item \textbf{Redirection:} Guide the interaction back to the session goal or toward appropriate external support when needed.
\end{itemize}

\subsection{Response Protocols by Content Category}
Application-level safety was specified through category-specific response protocols in the system prompts, each built on the shared principles above. Below we summarize the instructed response for each content category.

\textbf{Session Management.}
\begin{itemize}
    \item \textit{Ending session (technical):} Users asking how to end the session were instructed to press the red button with a square icon located at the bottom of the interface.
\end{itemize}

\textbf{High-Risk Safety Scenarios.}
\begin{itemize}
    \item \textit{Self-harm ideation:} ARC was instructed to avoid actionable guidance, clarify its non-human role, and direct users to external crisis resources (e.g., 988 Suicide \& Crisis Lifeline or 911 if in immediate danger). Responses were to acknowledge distress compassionately, and the protocol specified ending the session.
    \item \textit{Violent intent:} ARC was instructed not to provide instructions, to acknowledge emotional distress without endorsing harm, to clarify scope limitations, and to redirect to safe topics.
\end{itemize}

\textbf{Harmful or Boundary-Violating Content.}
\begin{itemize}
    \item \textit{Hate or offensive language:} ARC was instructed to avoid engaging with harmful content, reinforce dignity (e.g., affirming the value of older adults), clarify scope, and redirect.
    \item \textit{Sexual or romantic advances:} ARC was instructed to set boundaries, clarify its role, and redirect to session content.
\end{itemize}

\textbf{Health and Safety Requests.}
\begin{itemize}
    \item \textit{Medical advice requests:} ARC was instructed to decline diagnosis or treatment guidance, encourage consultation with healthcare professionals, and (when appropriate) support symptom reflection (e.g., frequency, duration, impact) without offering medical advice.
    \item \textit{False or unverified health claims:} ARC was instructed not to endorse misinformation, to encourage professional consultation, and to redirect.
    \item \textit{Dangerous activities:} ARC was instructed to decline to provide instructions, emphasize safety (e.g., “That sounds risky”), and redirect.
    \item \textit{Substance misuse:} ARC was instructed to avoid giving instructions, acknowledge distress, encourage professional support, and redirect.
\end{itemize}

\textbf{Vulnerability and Risk Contexts.}
\begin{itemize}
    \item \textit{Abuse or neglect:} ARC was instructed to avoid affirming self-blame, emphasize user safety and dignity, and encourage seeking trusted or professional support (e.g., emergency services if needed).
    \item \textit{Financial exploitation:} ARC was instructed to warn against sharing money or personal information, encourage verification with trusted individuals, and redirect.
\end{itemize}

\textbf{Privacy and System Boundaries.}
\begin{itemize}
    \item \textit{Requests for personal or system information:} ARC was instructed to decline disclosure and reinforce privacy and security boundaries.
    \item \textit{Technology and privacy concerns:} ARC was instructed to give simple explanations consistent with study consent (e.g., data storage, limited access, Azure safety monitoring), avoid false reassurance, and encourage contacting the research team.
    \item \textit{Role confusion:} ARC was instructed to clarify that it is an AI Health Assistant Tool (not a human, clinician, or caregiver) and redirect to appropriate support if needed.
\end{itemize}

\end{document}